\pdfoutput=1
\documentclass[prd,showpacs,letterpaper,10pt,aps,notitlepage,nofootinbib]{revtex4-2}

\usepackage[utf8]{inputenc}
\usepackage{amsfonts}
\usepackage{amsmath}
\usepackage{graphicx}
\usepackage{hyperref}
\usepackage{placeins}
\usepackage{mathrsfs}
\usepackage{bm}
\usepackage[utf8]{inputenc}
\usepackage{multirow}
\hypersetup{colorlinks, linkcolor = [rgb]{0,0.0,0.75}, citecolor = [rgb]{0,0.0,0.75}, urlcolor = [rgb]{0,0.0,0.75}}

\newcommand{\nb}{\phantom{0}}
\newcommand{\wm}{\phantom{-}}

\begin{document}

\title{\texorpdfstring{$\bm{\Lambda_c \to N}$}{Lambda_c to N} form factors from lattice QCD and \\
phenomenology of \texorpdfstring{$\bm{\Lambda_c \to n\, \ell^+ \nu_\ell}$}{Lambda_c to n l+ nu}
and \texorpdfstring{$\bm{\Lambda_c \to p\,\mu^+\mu^-}$}{Lambda_c to p mu+ mu-} decays}

\author{Stefan Meinel}
\affiliation{Department of Physics, University of Arizona, Tucson, AZ 85721, USA}
\affiliation{RIKEN BNL Research Center, Brookhaven National Laboratory, Upton, NY 11973, USA}


\begin{abstract}
A lattice QCD determination of the $\Lambda_c \to N$ vector, axial vector, and tensor form factors is reported.
The calculation was performed with $2+1$ flavors of domain wall fermions at lattice spacings of $a\approx 0.11\:{\rm fm},\:0.085\:{\rm fm}$
and pion masses in the range $230\:{\rm MeV} \lesssim m_\pi \lesssim 350$ MeV. The form factors are extrapolated to the continuum limit
and the physical pion mass using modified $z$ expansions. The rates of the charged-current decays $\Lambda_c \to n\, e^+ \nu_e$ and
$\Lambda_c \to n\, \mu^+ \nu_\mu$ are predicted to be $\left( 0.405 \pm 0.016_{\,\rm stat} \pm 0.020_{\,\rm syst} \right)|V_{cd}|^2 \:{\rm ps}^{-1}$
and $\left( 0.396 \pm 0.016_{\,\rm stat} \pm 0.020_{\,\rm syst} \right)|V_{cd}|^2 \:{\rm ps}^{-1}$, respectively. The phenomenology
of the rare charm decay $\Lambda_c \to p\,\mu^+\mu^-$ is also studied. The differential branching fraction, the fraction of longitudinally polarized dimuons, and
the forward-backward asymmetry are calculated in the Standard Model and in an illustrative new-physics scenario.
\end{abstract}

\maketitle

\FloatBarrier
\section{Introduction}
\FloatBarrier

This paper reports a lattice QCD calculation of the form factors describing the matrix elements $\langle N | \bar{q} \Gamma c|\Lambda_c\rangle$, where
$q$ denotes the up or down quark field, $N$ denotes the proton or neutron, and $\Gamma\in\left\{\gamma^\mu,\gamma^\mu\gamma_5,\sigma_{\mu\nu}\right\}$.
The calculation was done in the isospin-symmetric limit with $m_u=m_d$, in which the $\Lambda_c\to n$ and $\Lambda_c \to p$ form factors are exactly equal:
$\langle n | \bar{d}\Gamma c|\Lambda_c\rangle=\langle p | \bar{u}\Gamma c|\Lambda_c\rangle$.
The $\Lambda_c \to n$ vector and axial vector form factors govern the charged-current decays $\Lambda_c \to n\, \ell^+ \nu_\ell$,
whose rates are proportional to $|V_{cd}|^2$. While the combination of a neutron and a neutrino in the final state makes
measurements of these processes difficult, a precise first-principles calculation is still valuable, primarily to test other theoretical approaches
\cite{Ivanov:1996fj, Pervin:2005ve, Azizi:2009wn, Khodjamirian:2011jp, Gutsche:2014zna, Lu:2016ogy, Faustov:2016yza, Li:2016qai}.
The $\Lambda_c \to p$ form factors play a role in the rare charm decays $\Lambda_c \to p\,\ell^+\ell^-$, $\Lambda_c \to p\,\gamma$, and others.
Rare charm decays provide an opportunity to search for new fundamental physics, but this is more challenging than in the bottom sector due to the dominance of
long-distance contributions from nonlocal matrix elements in most or all of the kinematic range (except for some observables that vanish in the Standard Model).
Recent theoretical studies of mesonic rare charm decays such as $D^+ \to \pi^+\mu^+\mu^-$ can be found in Refs.~\cite{deBoer:2015boa, Fajfer:2015mia, Feldmann:2017izn}.
This work focuses on the decay $\Lambda_c \to p\,\mu^+\mu^-$, which was recently analyzed by the LHCb Collaboration \cite{Aaij:2017nsd}.
In the dimuon mass region excluding $\pm40$ MeV intervals around $m_\omega$ and $m_\phi$, an upper limit of
$\mathcal{B}(\Lambda_c \to p\,\mu^+\mu^-) < 7.7\times 10^{-8}$ at 90\% confidence level was obtained \cite{Aaij:2017nsd},
which is a substantial improvement over previous limits set by the BaBar \cite{Lees:2011hb} and Fermilab E653 \cite{Kodama:1995ia} Collaborations.

The remainder of the paper is structured as follows: The form factors are defined in Sec.~\ref{sec:FFdef}. The lattice QCD calculation is
described in Sec.~\ref{sec:lattice}. The predictions for the $\Lambda_c \to n\, e^+ \nu_e$ and $\Lambda_c \to n\, \mu^+ \nu_\mu$ decay rates
are presented in Sec.~\ref{sec:Lcn}, and the rare decay $\Lambda_c \to p\,\mu^+\mu^-$ is analyzed in Sec.~\ref{sec:Lcp}.
Conclusions are given in Sec.~\ref{sec:conclusions}.

\FloatBarrier
\section{\label{sec:FFdef}Definition of the form factors}
\FloatBarrier

In the following, we consider the proton final state for definiteness. This calculation uses the helicity-based definition of the form factors
introduced in Ref.~\cite{Feldmann:2011xf}, which is given by
\begin{eqnarray}
 \nonumber \langle N^+(p^\prime,s^\prime) | \overline{u} \,\gamma^\mu\, c | \Lambda_c(p,s) \rangle &=&
 \overline{u}_N(p^\prime,s^\prime) \bigg[ f_0(q^2)\: (m_{\Lambda_c}-m_N)\frac{q^\mu}{q^2} \\
 \nonumber && \phantom{\overline{u}_N \bigg[}+ f_+(q^2) \frac{m_{\Lambda_c}+m_N}{s_+}\left( p^\mu + p^{\prime \mu} - (m_{\Lambda_c}^2-m_N^2)\frac{q^\mu}{q^2}  \right) \\
 && \phantom{\overline{u}_N \bigg[}+ f_\perp(q^2) \left(\gamma^\mu - \frac{2m_N}{s_+} p^\mu - \frac{2 m_{\Lambda_c}}{s_+} p^{\prime \mu} \right) \bigg] u_{\Lambda_c}(p,s), \\
 \nonumber \langle N^+(p^\prime,s^\prime) | \overline{u} \,\gamma^\mu\gamma_5\, c | \Lambda_c(p,s) \rangle &=&
 -\overline{u}_N(p^\prime,s^\prime) \:\gamma_5 \bigg[ g_0(q^2)\: (m_{\Lambda_c}+m_N)\frac{q^\mu}{q^2} \\
 \nonumber && \phantom{\overline{u}_N \bigg[}+ g_+(q^2)\frac{m_{\Lambda_c}-m_N}{s_-}\left( p^\mu + p^{\prime \mu} - (m_{\Lambda_c}^2-m_N^2)\frac{q^\mu}{q^2}  \right) \\
 && \phantom{\overline{u}_N \bigg[}+ g_\perp(q^2) \left(\gamma^\mu + \frac{2m_N}{s_-} p^\mu - \frac{2 m_{\Lambda_c}}{s_-} p^{\prime \mu} \right) \bigg]  u_{\Lambda_c}(p,s),
\end{eqnarray}
\begin{eqnarray}
 \nonumber \langle N^+(p^\prime,s^\prime) | \overline{u} \,i\sigma^{\mu\nu} q_\nu \, c | \Lambda_c(p,s) \rangle &=&
 - \overline{u}_N(p^\prime,s^\prime) \bigg[  h_+(q^2) \frac{q^2}{s_+} \left( p^\mu + p^{\prime \mu} - (m_{\Lambda_c}^2-m_{\Lambda}^2)\frac{q^\mu}{q^2} \right) \\
 && \phantom{\overline{u}_N \bigg[} + h_\perp(q^2)\, (m_{\Lambda_c}+m_N) \left( \gamma^\mu -  \frac{2  m_N}{s_+} \, p^\mu - \frac{2m_{\Lambda_c}}{s_+} \, p^{\prime \mu}   \right) \bigg] u_{\Lambda_c}(p,s), \\
 \nonumber \langle N^+(p^\prime,s^\prime)| \overline{u} \, i\sigma^{\mu\nu}q_\nu \gamma_5  \, c|\Lambda_c(p,s)\rangle &=&
 -\overline{u}_{\Lambda}(p^\prime,s^\prime) \, \gamma_5 \bigg[   \widetilde{h}_+(q^2) \, \frac{q^2}{s_-} \left( p^\mu + p^{\prime \mu} -  (m_{\Lambda_c}^2-m_{\Lambda}^2) \frac{q^\mu}{q^2} \right) \\
 && \phantom{\overline{u}_N \bigg[}  + \widetilde{h}_\perp(q^2)\,  (m_{\Lambda_c}-m_N) \left( \gamma^\mu +  \frac{2 m_N}{s_-} \, p^\mu - \frac{2 m_{\Lambda_c}}{s_-} \, p^{\prime \mu}  \right) \bigg]  u_{\Lambda_c}(p,s),
\end{eqnarray}
where $q=p-p^\prime$, $\sigma^{\mu\nu}=\frac{i}{2}(\gamma^\mu\gamma^\nu-\gamma^\nu\gamma^\mu)$, and $s_\pm =(m_{\Lambda_c} \pm m_N)^2-q^2$.
These form factors satisfy the endpoint relations
\begin{eqnarray}
 f_0(0) &=& f_+(0), \label{eq:FFC1} \\
 g_0(0) &=& g_+(0), \label{eq:FFC2} \\
 g_\perp(q^2_{\rm max}) &=& g_+(q^2_{\rm max}), \label{eq:FFC3} \\
 \widetilde{h}_\perp(q^2_{\rm max}) &=& \widetilde{h}_+(q^2_{\rm max}), \label{eq:FFC4}
\end{eqnarray}
where $q^2_{\rm max}=(m_{\Lambda_c}-m_N)^2$.

\FloatBarrier
\section{\label{sec:lattice}Lattice calculation}
\FloatBarrier

\FloatBarrier
\subsection{\label{sec:latticeparams}Lattice parameters and correlation functions}
\FloatBarrier

This calculation uses the Iwasaki action \cite{Iwasaki:1984cj} for the gluons, the Shamir-type
domain-wall action \cite{Kaplan:1992bt, Furman:1994ky, Shamir:1993zy} for the $u$, $d$, and $s$ quarks,
and an anisotropic clover action for the $c$ quark, with parameters tuned in Ref.~\cite{Brown:2014ena} to reproduce the correct charmonium mass and relativistic
dispersion relation. The gauge field ensembles were generated by the RBC and UKQCD Collaborations and are described in detail in Ref.~\cite{Aoki:2010dy}.
This work is based on the same six sets of light-quark domain-wall propagators as Ref.~\cite{Detmold:2015aaa}; the parameters of these
sets and the resulting pion masses are listed in Table \ref{tab:params}. The parameters of the charm-quark action are given in the first
four rows of Table \ref{tab:paramsII}.

\begin{table}
\begin{tabular}{ccccccccccccccccccccc}
\hline\hline
Set & \hspace{1ex} & $\beta$ & \hspace{1ex} & $N_s^3\times N_t$ & \hspace{1ex} & $am_{u,d}^{(\mathrm{sea})}$
& \hspace{1ex} & $am_{s}^{(\mathrm{sea})}$   & \hspace{1ex} & $a$ [fm] & \hspace{1ex} & $am_{u,d}^{(\mathrm{val})}$ 
& \hspace{1ex} & $m_\pi^{(\mathrm{val})}$ [MeV]  & \hspace{1ex} & $N_{\rm samples}$ \\
\hline
\texttt{C14} && $2.13$ && $24^3\times64$ && $0.005$   && $0.04$   && $0.1119(17)$  && $0.001$   && 245(4) && 2672 \\
\texttt{C24} && $2.13$ && $24^3\times64$ && $0.005$   && $0.04$   && $0.1119(17)$  && $0.002$   && 270(4) && 2676 \\
\texttt{C54} && $2.13$ && $24^3\times64$ && $0.005$   && $0.04$   && $0.1119(17)$  && $0.005$   && 336(5) && 2782 \\
\texttt{F23} && $2.25$ && $32^3\times64$ && $0.004$   && $0.03$   && $0.0849(12)$  && $0.002$   && 227(3) && 1907 \\
\texttt{F43} && $2.25$ && $32^3\times64$ && $0.004$   && $0.03$   && $0.0849(12)$  && $0.004$   && 295(4) && 1917 \\
\texttt{F63} && $2.25$ && $32^3\times64$ && $0.006$   && $0.03$   && $0.0848(17)$  && $0.006$   && 352(7) && 2782 \\
\hline\hline
\end{tabular}
\caption{\label{tab:params}Parameters of the lattice gauge field ensembles \cite{Aoki:2010dy} and light quark propagators. The
\texttt{C14}/\texttt{C24} and \texttt{F23} data sets are based on the same gauge field ensembles as the \texttt{C54} and \texttt{F43}
data sets, respectively, and differ only in the valence quark mass used for the propagators.
The lattice spacing values given here were determined using the $\Upsilon(2S)-\Upsilon(1S)$ splitting in Ref.~\cite{Meinel:2010pv}.}
\end{table}

The $\Lambda_c$ and nucleon two-point functions and the $\Lambda_c \to N$ three-point functions were computed analogously to
Ref.~\cite{Detmold:2015aaa}, with the bottom quark replaced by the charm quark. The $c\to u$ currents were renormalized
using the ``mostly nonperturbative'' method \cite{Hashimoto:1999yp, ElKhadra:2001rv}, in which the bulk of the renormalization is absorbed by an overall factor of
$\sqrt{Z_V^{(uu)} Z_V^{(cc)}}$, where $Z_V^{(uu)}$ and $Z_V^{(cc)}$ are the nonperturbative renormalization factors of the
currents $\bar{u}\gamma_0 u$ and $\bar{c}\gamma_0 c$. The $c\to u$ vector and axial vector currents
are defined as in Eqs.~(18)-(21) of Ref.~\cite{Detmold:2015aaa}, with residual matching factors and $\mathcal{O}(a)$-improvement
coefficients computed to one loop by Christoph Lehner \cite{Lehner:2012bt, Lehnercharmlight} and listed in Table \ref{tab:paramsII}.
The full one-loop $\mathcal{O}(a)$ improvement was performed for the vector and axial vector
currents for all source-sink separations in the three-point functions (instead of just a subset of separations as in Ref.~\cite{Detmold:2015aaa}).
For the $c \to u$ tensor currents, the one-loop calculation was not available, and the perturbative coefficients were evaluated at
tree-level as in Ref.~\cite{Detmold:2016pkz}. That is, the tensor currents were written as
\begin{equation}
 T_{\mu\nu}=\sqrt{Z_V^{(uu)} Z_V^{(cc)}} \bigg[ \bar{u} \sigma_{\mu\nu} c + a\, d_1\, \sum_{j=1}^3\bar{u} \sigma_{\mu\nu} \gamma_j \overrightarrow{\nabla}_j  c \bigg],
\end{equation}
where $d_1$ is the mean-field-improved heavy-quark ``field rotation'' coefficient \cite{ElKhadra:1996mp}, whose values are
also listed in Table \ref{tab:paramsII}. The missing one-loop corrections result in larger systematic uncertainties for
the tensor form factors, as discussed in Sec.~\ref{sec:zexp}.

\begin{table}
\begin{center}
\small
\begin{tabular}{cllll}
\hline\hline
Parameter          & \hspace{2ex} & \hspace{1ex} Coarse lattice    & \hspace{2ex} &  \hspace{1ex} Fine lattice       \\
\hline
 $a m_c$                     && $\wm0.1214$           && $-0.0045$           \\[0.2ex]
 $\nu$                       && $\wm1.2362$           && $\wm1.1281$         \\[0.2ex]
 $c_{E}$                     && $\wm1.6650$           && $\wm1.5311$         \\[0.2ex]
 $c_{B}$                     && $\wm1.8409$           && $\wm1.6232$         \\[0.2ex]
 $d_1$                       && $\wm0.0437$           && $\wm0.0355$          \\[0.2ex]
 $Z_V^{(cc)}$                && $\wm1.35725(23)$      && $\wm1.18321(14)$    \\[0.2ex]   
 $Z_V^{(uu)}$                && $\wm0.71651(46)$      && $\wm0.74475(12)$    \\[0.2ex]
 $\rho_{V^0}=\rho_{A^0}$     && $\wm1.00274(49)$      && $\wm1.001949(85)$   \\[0.2ex]
 $\rho_{V^j}=\rho_{A^j}$     && $\wm0.99475(62)$      && $\wm0.99675(68)$    \\[0.2ex]
 $c_{V^0}^R=c_{A^0}^R$       && $\wm0.0402(88)$       && $\wm0.0353(92)$     \\[0.2ex]
 $c_{V^0}^L=c_{A^0}^L$       && $-0.0048(48)$         && $-0.0027(28)$       \\[0.2ex]
 $c_{V^j}^R=c_{A^j}^R$       && $\wm0.0346(51)$       && $\wm0.0283(43)$     \\[0.2ex]
 $c_{V^j}^L=c_{A^j}^L$       && $\wm0.00012(26)$      && $\wm0.00040(42)$    \\[0.2ex]
 $d_{V^j}^R=-d_{A^j}^R$      && $-0.0041(41)$         && $-0.0039(39)$       \\[0.2ex]
 $d_{V^j}^L=-d_{A^j}^L$      && $\wm0.0021(21)$       && $\wm0.0026(26)$     \\[0.2ex]
\hline\hline
\end{tabular}\vspace{-2ex}
\end{center}
\caption{\label{tab:paramsII} Charm-quark action parameters \cite{Brown:2014ena}
and $c\to u$ current matching and $\mathcal{O}(a)$ improvement parameters \cite{Lehner:2012bt, Lehnercharmlight}.
The nonperturbative factors $Z_V^{(uu)}$ and $Z_V^{(cc)}$ were determined in Refs.~\cite{Blum:2014tka} and \cite{Detmold:2015aaa}, respectively.
The uncertainties of the residual matching factors and $\mathcal{O}(a)$-improvement coefficients were estimated as
$(h^{(0)} / h^{(1)} - 1)^2 h^{(1)}$, where $h^{(0)}$ is the tree-level result and $h^{(1)}$ is the full one-loop result \cite{Lehner:2012bt, Lehnercharmlight}.}
\end{table}

The three-point functions were computed for all source-sink separations in the ranges
$t/a=4...15$ (\texttt{C14}, \texttt{C24}, \texttt{C54} data sets), $t/a=5...15$ (\texttt{\texttt{F43}} data set),
and $t/a=5...17$ (\texttt{\texttt{F63}} data set). The $\Lambda_c$ momentum, $\mathbf{p}$, was set to zero,
and all nucleon momenta $\mathbf{p}^\prime$ with $1\left(\frac{2\pi}{L}\right)^2 \leq|\mathbf{p}^\prime|^2\leq 5\left(\frac{2\pi}{L}\right)^2$
were used. From the three-point and two-point correlation functions, the ``ratios''
$R_{f_\perp}(|\mathbf{p}^\prime|, t)$, $R_{f_+}(|\mathbf{p}^\prime|, t)$, $R_{f_0}(|\mathbf{p}^\prime|, t)$,
$R_{g_\perp}(|\mathbf{p}^\prime|, t)$, $R_{g_+}(|\mathbf{p}^\prime|, t)$, $R_{g_0}(|\mathbf{p}^\prime|, t)$,
$R_{h_\perp}(|\mathbf{p}^\prime|, t)$, $R_{h_+}(|\mathbf{p}^\prime|, t)$, $R_{\widetilde{h}_\perp}(|\mathbf{p}^\prime|, t)$,
and $R_{\widetilde{h}_+}(|\mathbf{p}^\prime|, t)$, defined as in Eqs.~(52-54), (58-60) of Ref.~\cite{Detmold:2015aaa}
and Eqs.~(27-30) of Ref.~\cite{Detmold:2016pkz} (with the appropriate replacements of initial and final baryons), were computed. These ratios are equal to
the form factors $f_\perp$, $f_+$, ... for the given momentum and lattice parameters, up to
excited-state contributions that vanish exponentially as $t\to\infty$. The ratios also depend on the baryon masses,
which were obtained by fitting the two-point functions from the same data sets and are listed in Table \ref{tab:hadronmasses}
(the table also contains the $D$ meson masses, which are needed at a later stage in the analysis).

\begin{table}
\begin{tabular}{ccccccc}
\hline\hline
Set  & \hspace{1ex} & $a m_{\Lambda_c}$ & \hspace{1ex} & $am_N$  & \hspace{1ex} & $am_{D}$     \\
\hline
\texttt{C14}       &&  $1.3499(51)\nb$  &&  $0.6184(76)$  &&  $1.0728(12)\nb$  \\
\texttt{C24}       &&  $1.3526(57)\nb$  &&  $0.6259(57)$  &&  $1.0713(14)\nb$  \\
\texttt{C54}       &&  $1.3706(40)\nb$  &&  $0.6580(39)$  &&  $1.0763(13)\nb$  \\ 
\texttt{F23}       &&  $1.008(12)\nb$   &&  $0.4510(86)$  &&  $0.8092(11)\nb$  \\ 
\texttt{F43}       &&  $1.0185(67)\nb$  &&  $0.4705(42)$  &&  $0.81185(91)$    \\ 
\texttt{F63}       &&  $1.0314(40)\nb$  &&  $0.5004(25)$  &&  $0.81722(56)$    \\ 
\hline\hline
\end{tabular}
\caption{\label{tab:hadronmasses}Hadron masses in lattice units.}
\end{table}

\begin{figure}
 \includegraphics[width=0.495\linewidth]{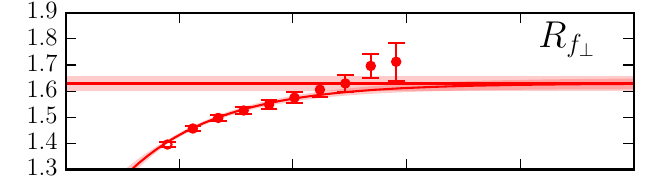} \hfill \includegraphics[width=0.495\linewidth]{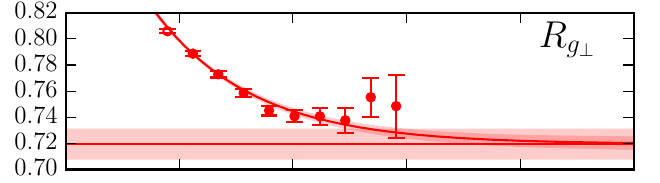}  \\
 \includegraphics[width=0.495\linewidth]{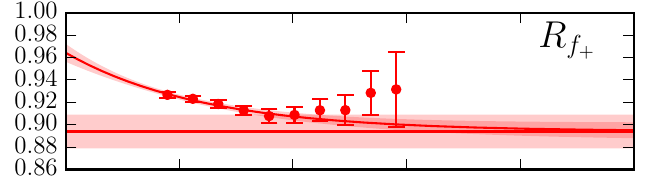} \hfill \includegraphics[width=0.495\linewidth]{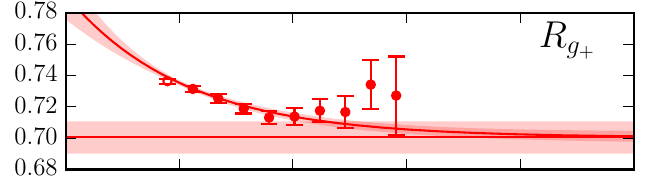}  \\
 \includegraphics[width=0.495\linewidth]{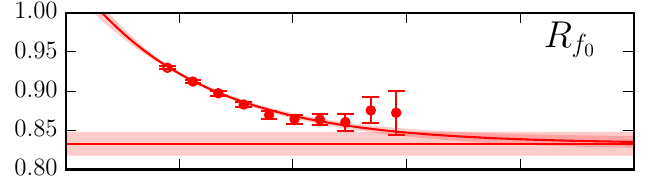} \hfill \includegraphics[width=0.495\linewidth]{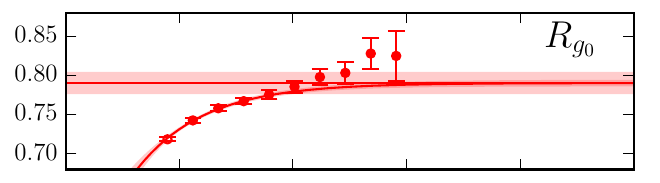}  \\
 \includegraphics[width=0.495\linewidth]{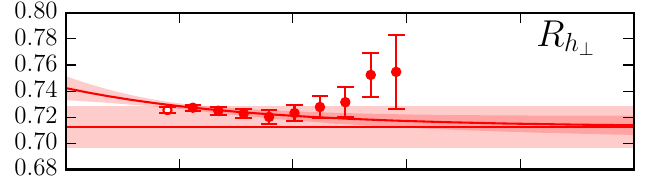} \hfill \includegraphics[width=0.495\linewidth]{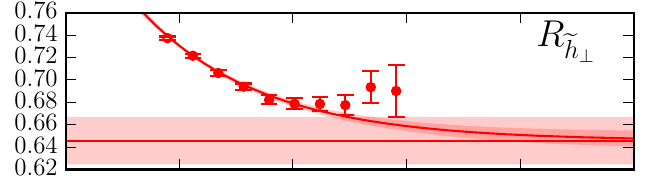}  \\
 \includegraphics[width=0.495\linewidth]{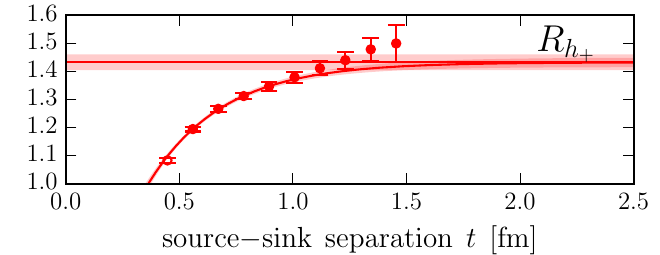} \hfill \includegraphics[width=0.495\linewidth]{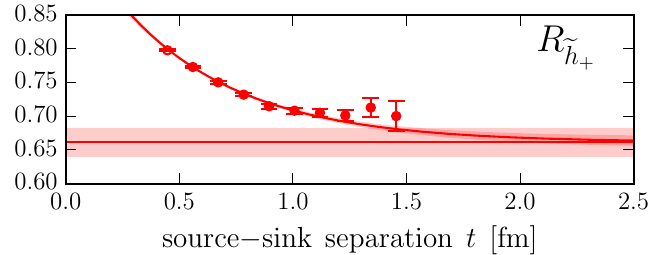}  \\
 
 \caption{\label{fig:tsepextrap}Fits of the quantities $R_{f}(|\mathbf{p}^\prime|, t)$ that are used to extract the ground-state
 form factors, $f(|\mathbf{p}^\prime|)=\lim_{t\to\infty}R_{f}(|\mathbf{p}^\prime|, t)$. The data shown here are from the \texttt{C54} set at
 $|\mathbf{p}^\prime|^2=2\left(\frac{2\pi}{L}\right)^2$. At a given momentum, the fits were performed jointly for all six data sets
 with constraints as explained in Ref.~\cite{Detmold:2015aaa}. Some data points at the shortest source-sink separations, plotted
 with open symbols, were excluded from the fits to achieve $\chi^2/{\rm d.o.f.}\lesssim 1$. The curves going through the data points
 show the fit functions of the form $R_f(|\mathbf{p}^\prime|,t)=f(|\mathbf{p}^\prime|)+A_f(|\mathbf{p}^\prime|)\, e^{-\delta_f(|\mathbf{p}^\prime|)\,t}$,
 with bands giving the statistical uncertainties. The horizonal lines show the extrapolated values $f(|\mathbf{p}^\prime|)$,
 where the bands also include estimates of the systematic uncertainties associated with the choices of fit ranges.}
\end{figure}

The ground-state form factors $f(|\mathbf{p}^\prime|)$ were then extracted by performing correlated fits of the ratios of the form
$R_f(|\mathbf{p}^\prime|,t)=f(|\mathbf{p}^\prime|)+A_f(|\mathbf{p}^\prime|)\, e^{-\delta_f(|\mathbf{p}^\prime|)\,t}$, which
include the leading excited-state contributions. Examples of the fits are shown in Fig.~\ref{fig:tsepextrap}.
At a given momentum, the fits were performed jointly for all six data sets, and jointly for the form factors associated
with a given type of current, with constraints as explained in Ref.~\cite{Detmold:2015aaa}. The constraints limit the variation
of the energy gap parameters across data sets to physically reasonable values, and enforce that the relations between the form factors
in the helicity basis and the ``Weinberg basis'' are preserved by the extrapolation \cite{Detmold:2015aaa,Detmold:2016pkz}.

The values of the start time slices $t_{\rm min}$ were chosen to achieve $\chi^2/{\rm d.o.f.}\lesssim 1$. The average of the
$\chi^2/{\rm d.o.f.}$ values of the twenty independent fits (four types of currents times five momenta) was 0.98, with a standard deviation of 0.16.
The number of degrees of freedom (d.o.f.), defined as the number of data points minus the number of unconstrained fit parameters, ranged from 94 to 294.
The average and standard deviation of the magnitudes $|A_f(|\mathbf{p}^\prime|)|$ of the excited-state-overlap parameters were $0.56 \pm 0.73$ for the form factors in the helicity basis
and $0.78 \pm 1.05$ for the form factors in the Weinberg basis.

To estimate the remaining systematic uncertainties associated with the choices of fit ranges,
additional fits were performed in which all $t_{\rm min}$ values were increased simultaneously by one unit.
As in Refs.~\cite{Detmold:2015aaa, Detmold:2016pkz, Meinel:2016dqj}, the systematic uncertainty in $f(|\mathbf{p}^\prime|)$ for a given momentum $|\mathbf{p}^\prime|$ and data set was estimated
as the larger of the following two: i) the shift in $f(|\mathbf{p}^\prime|)$ at the given momentum, and ii) the average of the shifts
in $f(|\mathbf{p}^\prime|)$ over all momenta. These systematic uncertainties were added to the statistical uncertainties in quadrature.
The results for $f(|\mathbf{p}^\prime|)$ with the combined uncertainties are listed in Table \ref{tab:FFlat} in the Appendix,
and are also shown as the horizontal bands in Fig.~\ref{fig:tsepextrap} and as the data points in Figs.~\ref{fig:V}-\ref{fig:T}.

\FloatBarrier
\subsection{\label{sec:zexp}Chiral and continuum extrapolations of the form factors}
\FloatBarrier

Following the extraction of the form factor values for each lattice data set and for each discrete momentum, global fits
of the form factor shape and of the dependence on the lattice spacing and quark masses were performed using modified BCL $z$-expansions
\cite{Bourrely:2008za}.
In the physical limit ($a=0$, $m_\pi=m_{\pi,{\rm phys}}$), the fit functions reduce to the form
\begin{equation}
 f(q^2) = \frac{1}{1-q^2/(m_{\rm pole}^f)^2} \sum_{n=0}^{n_{\rm max}} a_n^f [z(q^2)]^n, \label{eq:zexp}
\end{equation}
where the expansion variable is defined as
\begin{equation}
z(q^2) = \frac{\sqrt{t_+-q^2}-\sqrt{t_+-t_0}}{\sqrt{t_+-q^2}+\sqrt{t_+-t_0}}.
\end{equation}
This definition maps the complex $q^2$ plane, cut along the real axis for $q^2\geq t_+$, onto the disk $|z|<1$.
Here, $t_+$ is set equal to the threshold of $D\, \pi$ two-particle states,
\begin{equation}
t_+ = (m_D + m_\pi)^2. \label{eq:tplus}
\end{equation}
The parameter $t_0$ determines which value of $q^2$ gets mapped to $z=0$; in this work,
\begin{equation}
t_0 = q^2_{\rm max} = (m_{\Lambda_c} - m_N)^2
\end{equation}
is used. Furthermore, in Eq.~(\ref{eq:zexp}), the lowest poles are factored out before the $z$ expansion.
The quantum numbers and masses of the $D$ mesons producing these poles in the different form factors are given in Table \ref{tab:polemasses}.

To fit the lattice data, Eq.~(\ref{eq:zexp}) was augmented with additional parameters to describe the dependence on the lattice spacing
and on $m_\pi^2$ (which serves as a proxy for the the $u/d$ quark mass). As in Refs.~\cite{Detmold:2015aaa,Detmold:2016pkz},
two independent fits were performed: a ``nominal'' fit, which provides the central values and statistical uncertainties of the form factors,
and a ``higher-order (HO)'' fit that is used to estimate systematic uncertainties. In this work, the functions used for the nominal fit were
\begin{eqnarray}
\nonumber f(q^2) &=& \frac{1}{1-(a^2 q^2)/(a m_D + a \Delta^f)^2} \bigg[ a_0^f\bigg(1+c_0^f \frac{m_\pi^2-m_{\pi,{\rm phys}}^2}{\Lambda_\chi^2} \bigg) + a_1^f\:z(q^2) + a_2^f\:[z(q^2)]^2  \bigg] \\
 & & \times \bigg[1  + b^f\, a^2 |\mathbf{p^\prime}|^2 + d^f\, a^2 \Lambda_{\rm had}^2 \bigg],  \label{eq:FFccfit}
\end{eqnarray}
with free parameters $a_0^f$, $a_1^f$, $a_2^f$, $c_0^f$, $b^f$, and $d^f$. Here, the scales $\Lambda_\chi = 4\pi f_\pi$  with $f_\pi = 132\:\,{\rm MeV}$
and $\Lambda_{\rm had}=300\:\,{\rm MeV}$ were introduced to make all parameters dimensionless. The momentum transfers in lattice units, $a^2 q^2$,
were evaluated using the lattice results for the baryon masses from each data set, and their uncertainties and correlations were taken into account.
In addition, the pole masses were rewritten as $a m_{\rm pole}^f = a m_D + a \Delta^f$, where $\Delta^f$ are the mass differences relative to the
pseudoscalar $D$ meson. These mass differences were fixed to their physical values according to Table \ref{tab:polemasses}, while the lattice results
from each data set were used to evaluate $a m_D$.

The systematic uncertainties associated with the choices of $t_{\rm min}$ in the extractions of the lattice form factors from the ratios of correlation functions
(cf.~Sec.~\ref{sec:latticeparams}) were considered to be part of the statistical uncertainties in the $z$-expansion fits discussed here, and were therefore included
in the data covariance matrix for both the nominal and higher-order fits. Given that the procedure for estimating these uncertainties
was based on the magnitudes in the shifts, and conservatively used the larger of two choices, there is no obviously correct way of evaluating their correlations.
These systematic uncertainties were therefore added to the diagonal elements of the covariance matrices only.
As a result, the $z$-expansion fits have rather low values of $\chi^2/{\rm d.o.f.}$ (0.20 for the nominal fit and 0.19 for the higher-order fit, with ${\rm d.o.f.}=235$).

The functions used for the higher-order fit were
\begin{eqnarray}
\nonumber f_{\rm HO}(q^2) &=& \frac{1}{1-(a^2 q^2)/(a m_D + a \Delta^f)^2}
\bigg[ a_0^f\bigg(1+c_0^f \frac{m_\pi^2-m_{\pi,{\rm phys}}^2}{\Lambda_\chi^2}+\widetilde{c}_0^f \frac{m_\pi^3-m_{\pi,{\rm phys}}^3}{\Lambda_\chi^3}\bigg) \\
\nonumber && \hspace{30ex} +\: a_1^f\bigg(1+c_1^f\frac{m_\pi^2-m_{\pi,{\rm phys}}^2}{\Lambda_\chi^2}\bigg)\:z(q^2)  + a_2^f\:[z(q^2)]^2  + a_3^f\:[z(q^2)]^3 \bigg] \\
&& \times\: \bigg[1  + b^f\, a^2|\mathbf{p^\prime}|^2 + d^f\, a^2\Lambda_{\rm had}^2
                     + \widetilde{b}^f\, a^4 |\mathbf{p^\prime}|^4
                     + \widehat{d}^f\, a^3 \Lambda_{\rm had}^3
                     + \widetilde{d}^f a^4 \Lambda_{\rm had}^4
                     + j^f   a^4 |\mathbf{p^\prime}|^2\Lambda_{\rm had}^2 \bigg]. \hspace{5ex}  \label{eq:FFccfitHO}
\end{eqnarray}
The nominal fit already provides a good description of the lattice data, and the additional terms in the higher-order fit are used only
to estimate systematic uncertainties in a Bayesian approach. To this end, the additional parameters were constrained with Gaussian priors to be no larger than natural-sized.
The priors for the parameters $\widetilde{c}_0^f$, $c_1^f$, $\widetilde{b}^f$, $\widehat{d}^f$, $\widetilde{d}^f$, and $j^f$ were
chosen as in Ref.~\cite{Detmold:2016pkz}, while the $a_3^f$ were constrained to be $0\pm30$ as in Ref.~\cite{Meinel:2016dqj}. In the higher-order fit,
additional sources of systematic uncertainties were simultaneously incorporated as follows:
\begin{enumerate}
 \item When generating the bootstrap samples for the ratios $R_{f_\perp}(|\mathbf{p}^\prime|, t)$, $R_{f_+}(|\mathbf{p}^\prime|, t)$,
 $R_{f_0}(|\mathbf{p}^\prime|, t)$, $R_{g_\perp}(|\mathbf{p}^\prime|, t)$, $R_{g_+}(|\mathbf{p}^\prime|, t)$, $R_{g_0}(|\mathbf{p}^\prime|, t)$,
 the residual matching factors and $\mathcal{O}(a)$-improvement coefficients were drawn from Gaussian random distributions with central values
 and widths according to Table \ref{tab:paramsII}.
 \item To incorporate the systematic uncertainty in the tensor form factors due to the use of the tree-level values $\rho_{T^{\mu\nu}}=1$ for the
 residual matching factors, nuisance parameters $\rho_{T^{\mu\nu}}$ multiplying these form factors with Gaussian priors $1\pm\sigma_{\rho_{T^{\mu\nu}}}$
 were introduced in the fit. For the $b\to s$ currents in Ref.~\cite{Detmold:2016pkz}, $\sigma_{\rho_{T^{\mu\nu}}}$ was estimated to be equal to 2 times
 the maximum value of $|\rho_{V^{\mu}}-1|$, $|\rho_{A^{\mu}}-1|$, which was $0.05316$, comparable in magnitude to actual one-loop results for
 $|\rho_{T^{\mu\nu}}-1|$ obtained for staggered light quarks in Ref.~\cite{Bailey:2015dka}. For the $c\to u$ currents in this work,
 $\rho_{V^{\mu}}$ and $\rho_{A^{\mu}}$ are much closer to 1 (see Table \ref{tab:paramsII}), and the analogous procedure
 would yield $\sigma_{\rho_{T^{\mu\nu}}}=0.0105$. Given that the tensor current is scale-dependent and may exhibit qualitatively different
 behavior in a matching calculation (compared to the vector and axial vector currents), this appears to be too aggressive. The uncertainty
 estimate was therefore increased to $\sigma_{\rho_{T^{\mu\nu}}}=0.05$, which is approximately 10 times the maximum value of
 $|\rho_{V^{\mu}}-1|$, $|\rho_{A^{\mu}}-1|$. This estimate of the uncertainty (and the numerical values of the tensor form factors)
 should be understood as corresponding to the scale $\mu=m_c$.
 \item The finite-volume errors in the $\Lambda_c \to N$ form factors are expected to be approximately equal to those in the $\Lambda_b \to N$ form factors,
 which were estimated to be 3\% for the parameters of this calculation \cite{Detmold:2015aaa}. The missing isospin breaking effects are expected to be
 of order $\mathcal{O}((m_d-m_u)/\Lambda_{\rm QCD})\approx 0.5\%$ and $\mathcal{O}(\alpha_{\rm e.m.})\approx 0.7\%$. The uncertainties from these sources
 were added to the data correlation matrix used in the fit.
 \item The lattice spacings and pion masses of the different data sets were promoted to fit parameters, with Gaussian priors chosen according
 to their known central values and uncertainties.
\end{enumerate}

\begin{figure}
 \includegraphics[width=0.9\linewidth]{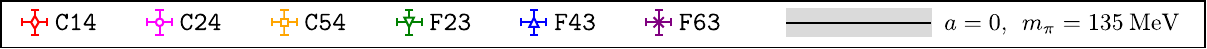}
 
 \vspace{1ex} 
 
 \includegraphics[width=\linewidth]{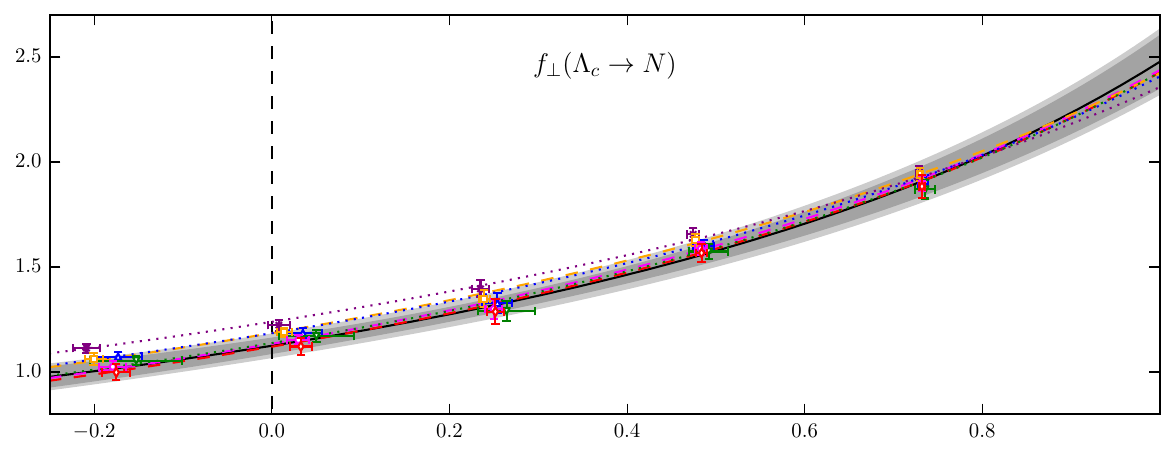} \\
 \includegraphics[width=\linewidth]{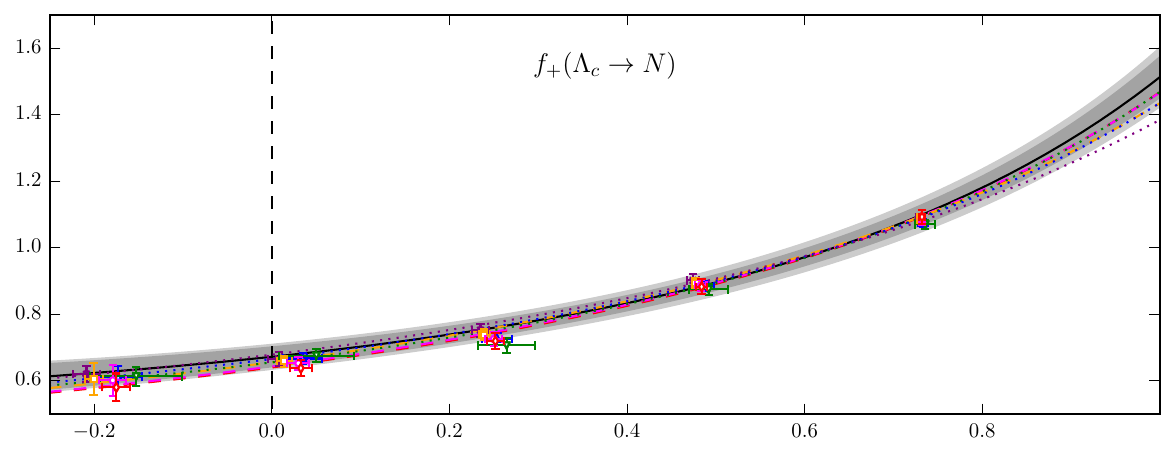} \\
 \includegraphics[width=\linewidth]{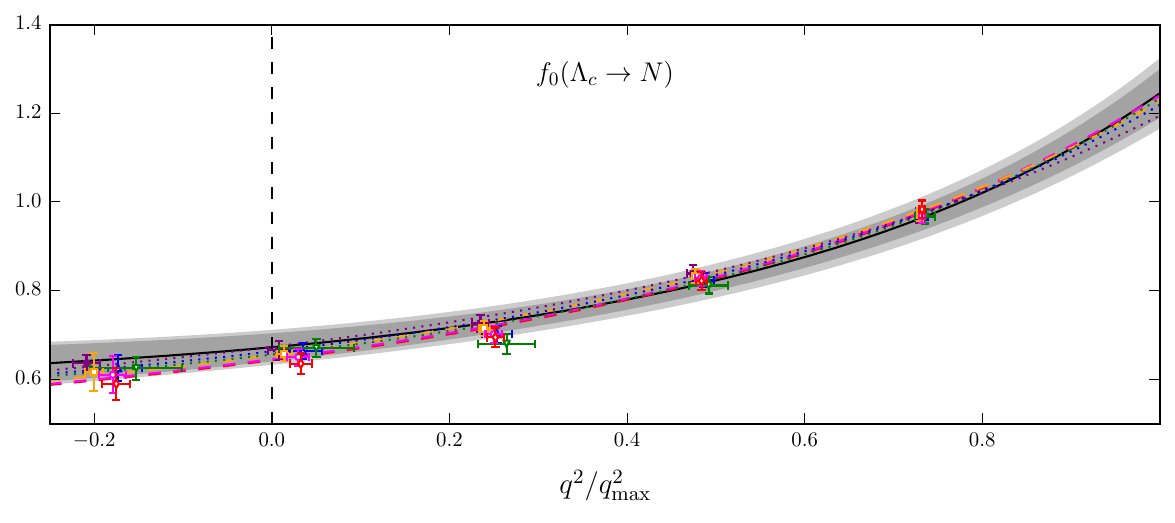}
 
 \caption{\label{fig:V}Lattice QCD results for the $\Lambda_c\to N$ vector form factors from the six different data sets (the labels are explained
 in Table \protect\ref{tab:params}), along with the modified $z$-expansion fits evaluated at the lattice parameters (dashed lines at $a\approx 0.11$ fm
 and dotted lines at $a\approx0.085$ fm) and in the physical limit (solid lines, with statistical and total uncertainties indicated by the inner and outer bands).}
\end{figure}

\begin{figure}
 \includegraphics[width=0.9\linewidth]{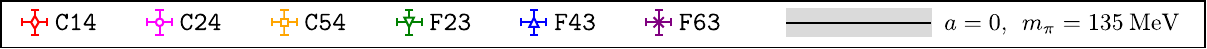}
 
 \vspace{1ex} 
 
 \includegraphics[width=\linewidth]{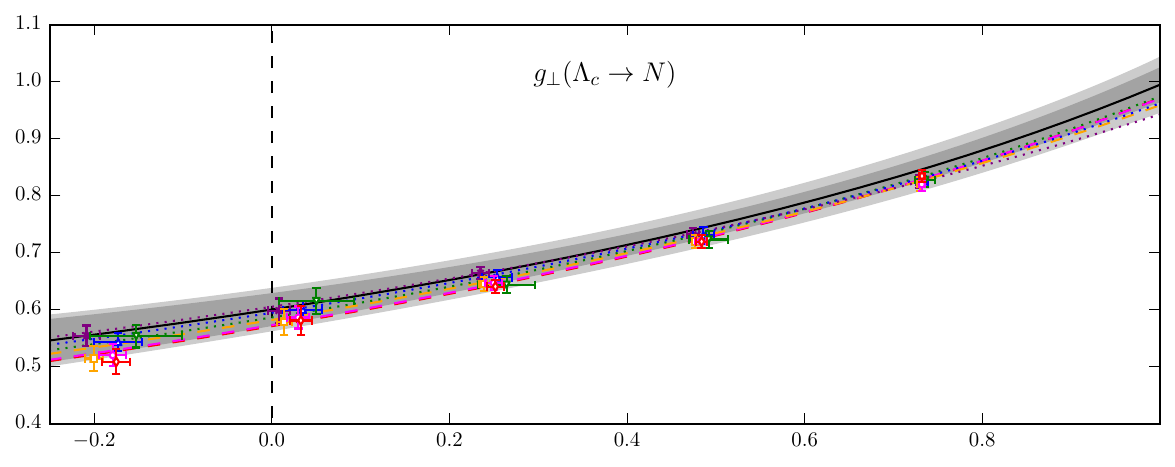} \\
 \includegraphics[width=\linewidth]{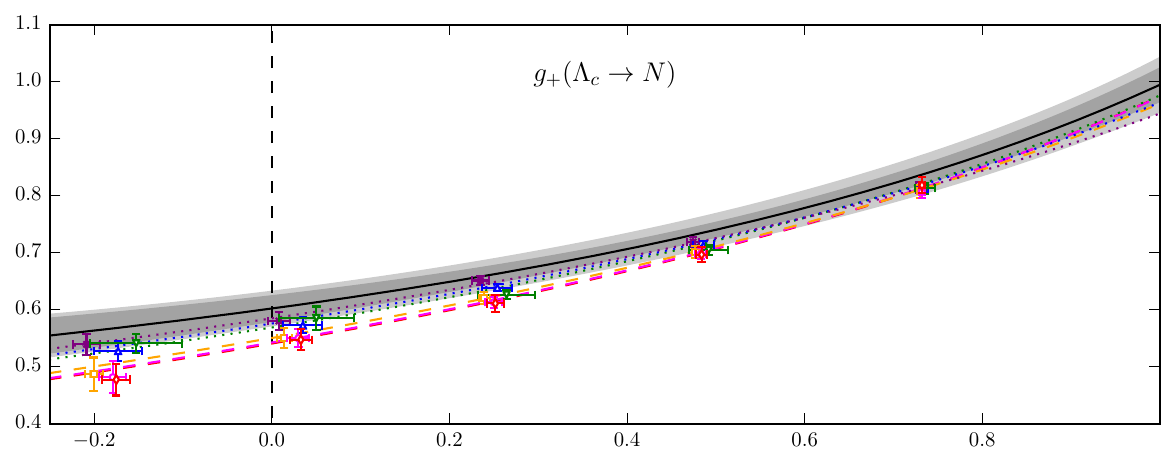} \\
 \includegraphics[width=\linewidth]{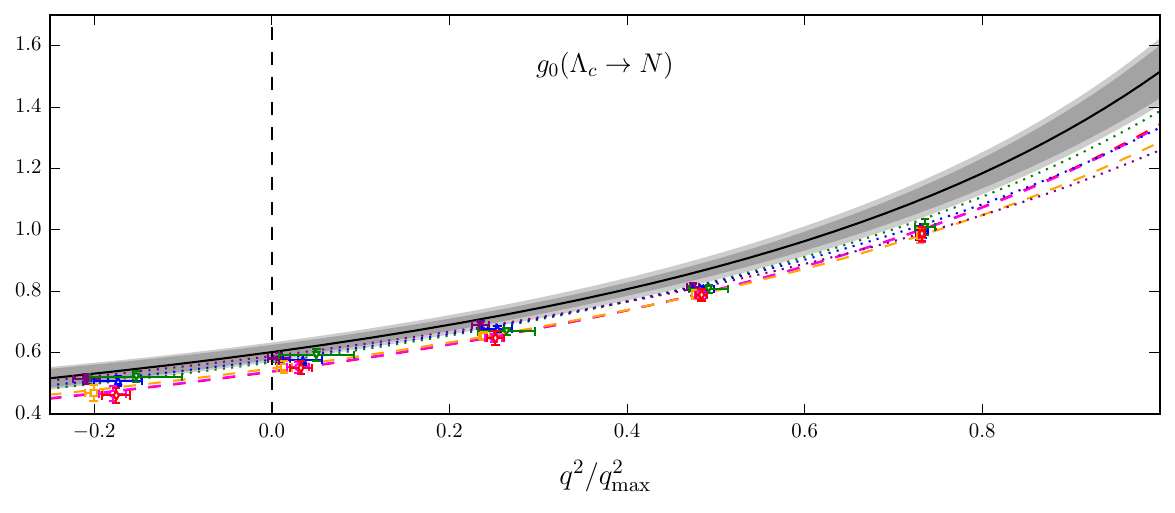}
 
 \caption{\label{fig:A}Like Fig.~\protect\ref{fig:V}, for the axial vector form factors.}
\end{figure}

\begin{figure}
 \includegraphics[width=0.9\linewidth]{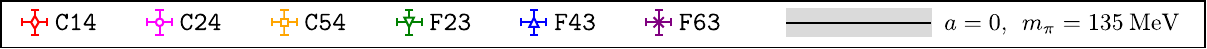}
 
 \vspace{1ex} 
 
 \includegraphics[width=\linewidth]{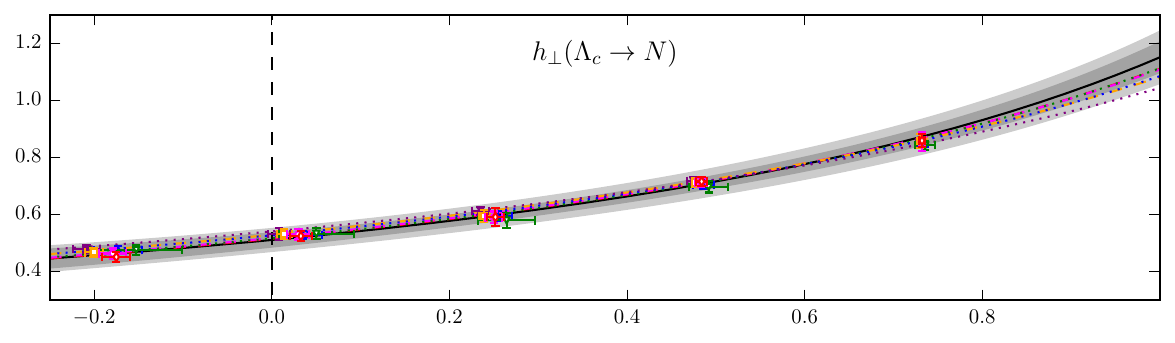} \\
 \includegraphics[width=\linewidth]{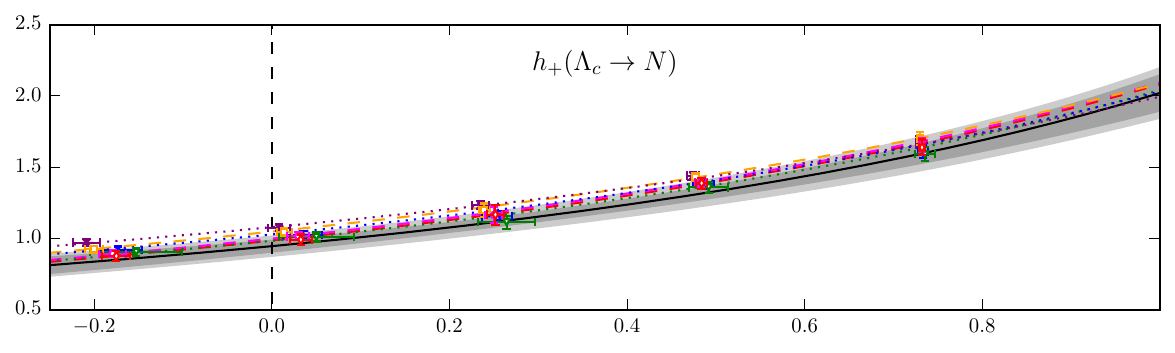} \\
 \includegraphics[width=\linewidth]{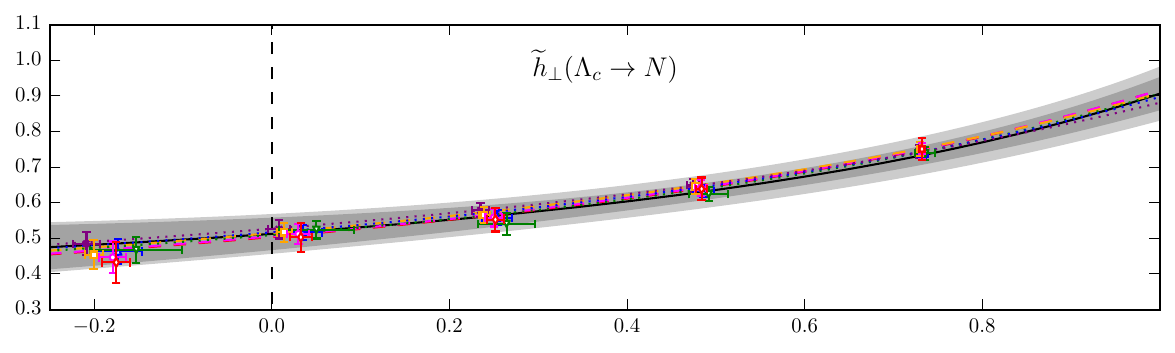} \\
 \includegraphics[width=\linewidth]{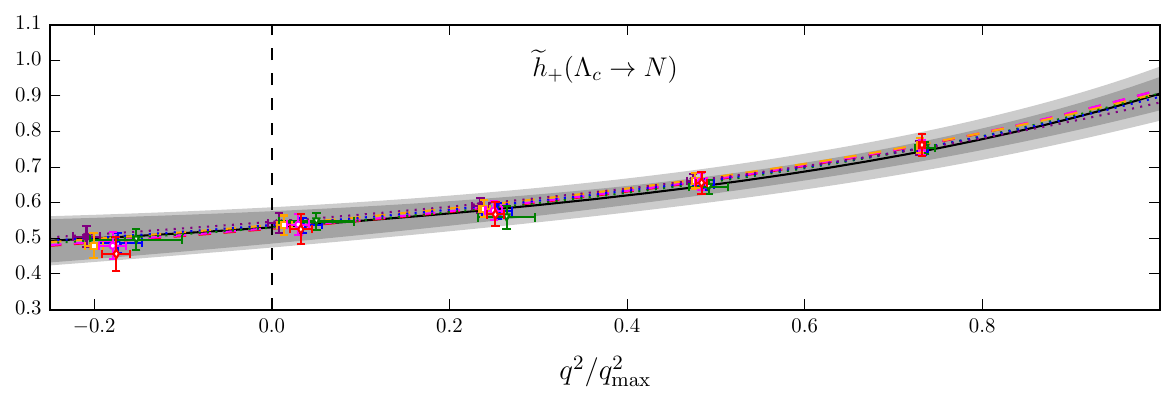} \\
 
 \caption{\label{fig:T}Like Fig.~\protect\ref{fig:V}, for the tensor form factors.}
\end{figure}

\noindent In the physical limit, the nominal and higher-order fits reduce to the form given in Eq.~(\ref{eq:zexp}),
with $n_{\rm max}=2$ and $n_{\rm max}=3$, respectively. The solid curves in Figs.~\ref{fig:V}, \ref{fig:A}, and \ref{fig:T}
with shaded error bands show the form factors in the physical limit. The results for the relevant parameters are given in Table
\ref{tab:fitresults}. Files containing the parameter values and the full covariance matrices are provided as supplemental material \cite{supplementalmaterial}.
The systematic uncertainty of any quantity depending on the form factors is estimated as
\begin{equation}
 \sigma_{O,{\rm syst}} = {\rm max}\left( |O_{\rm HO}-O|,\: \sqrt{|\sigma_{O,{\rm HO}}^2-\sigma_O^2|}  \right), \label{eq:sigmasyst}
\end{equation}
where $O$ and $O_{\rm HO}$ are the central values obtained from the nominal and higher-order parameterizations, and $\sigma_O$ and $\sigma_{O,{\rm HO}}$ are the
uncertainties propagated using the covariance matrices given in the supplemental material for the nominal and higher-order fit parameters.
See also Ref.~\cite{Detmold:2016pkz} for further explanations of this procedure.

\begin{figure}
 \hspace{1cm}\includegraphics[width=\linewidth]{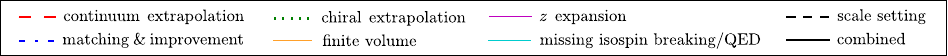}
 
 \vspace{0.5ex}
 
 \includegraphics[width=0.48\linewidth]{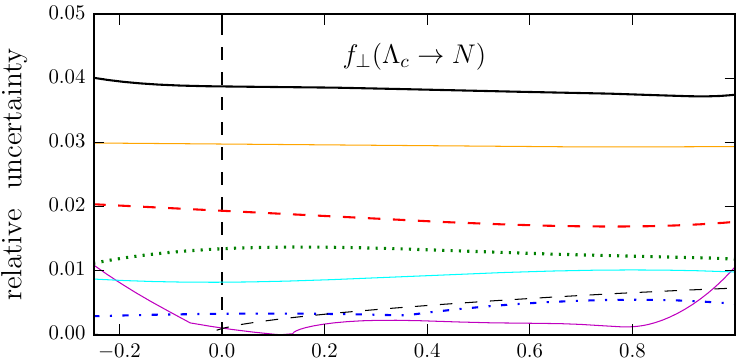} \hfill
 \includegraphics[width=0.48\linewidth]{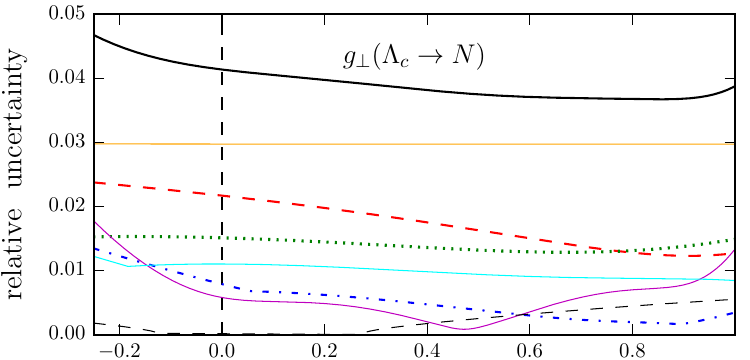} \\
 \includegraphics[width=0.48\linewidth]{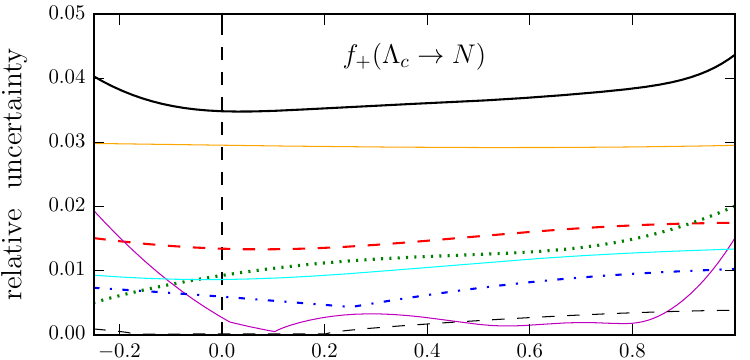} \hfill
 \includegraphics[width=0.48\linewidth]{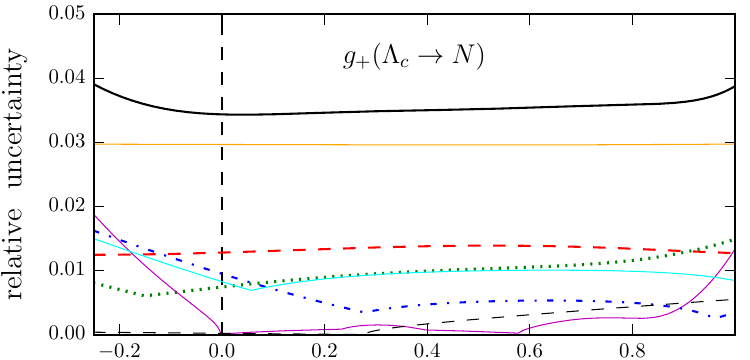} \\
 \includegraphics[width=0.48\linewidth]{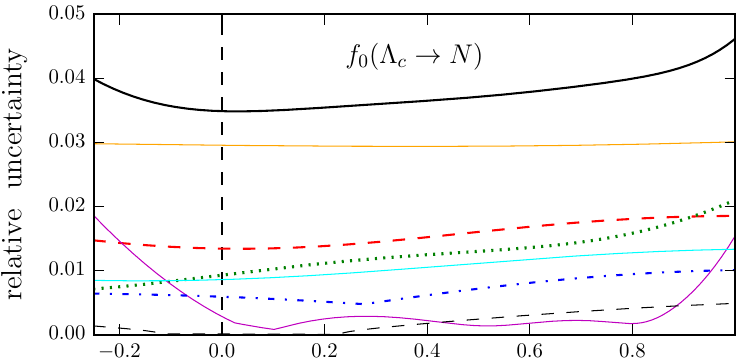} \hfill
 \includegraphics[width=0.48\linewidth]{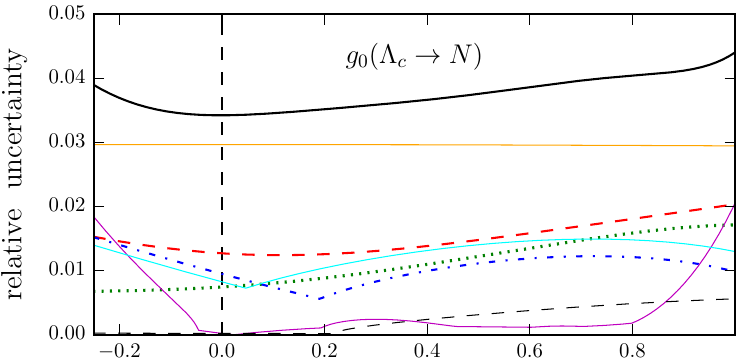} \\
 \includegraphics[width=0.48\linewidth]{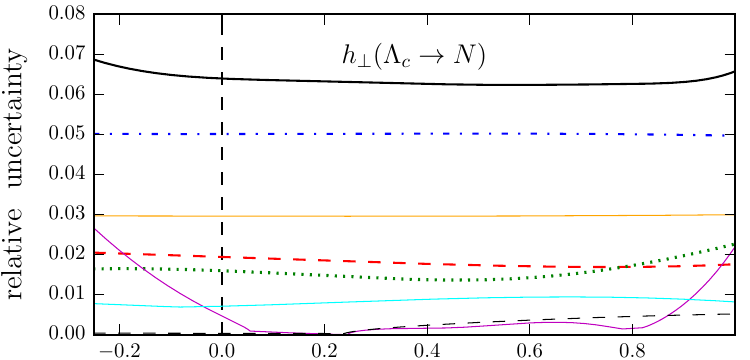} \hfill
 \includegraphics[width=0.48\linewidth]{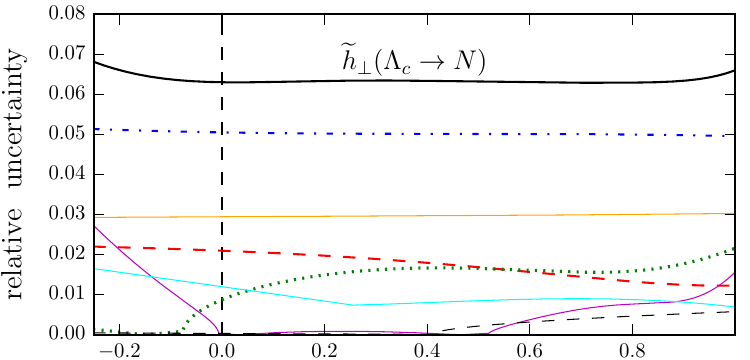} \\
 \includegraphics[width=0.48\linewidth]{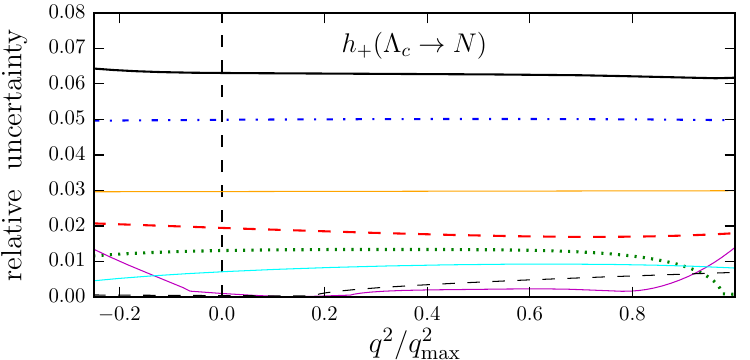} \hfill
 \includegraphics[width=0.48\linewidth]{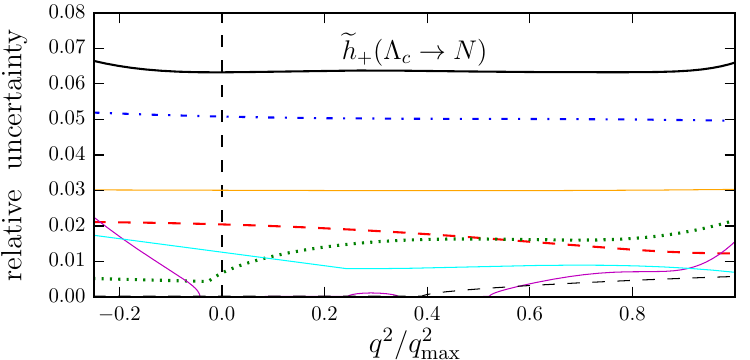}
 
 \caption{\label{fig:finalFFssyst}Breakdown of systematic uncertainties in the form factors.
 The individual uncertainties are shown for illustration only; the combined systematic uncertainty is evaluated directly using
 the full higher-order fit.}
\end{figure}

A breakdown of the form factor systematic uncertainties into individual sources is shown in Fig.~\ref{fig:finalFFssyst}.
This was obtained by performing additional fits to the lattice data where each one of the above modifications to the
fit functions or data covariance matrix was done individually, and then comparing each of these fits to the nominal fit as in Eq.~(\ref{eq:sigmasyst}).

\begin{table}
\begin{tabular}{ccccc}
\hline\hline
 $f$     & \hspace{1ex} & $J^P$ & \hspace{1ex}  & $m_{\rm pole}^f$ [GeV]  \\
\hline
$f_+$, $f_\perp$, $h_+$, $h_\perp$                         && $1^-$   && $2.010$  \\
$f_0$                                                      && $0^+$   && $2.351$  \\
$g_+$, $g_\perp$, $\widetilde{h}_+$, $\widetilde{h}_\perp$ && $1^+$   && $2.423$  \\
$g_0$                                                      && $0^-$   && $1.870$  \\
\hline\hline
\end{tabular}
\caption{\label{tab:polemasses} The quantum numbers and masses of the $D$ mesons producing poles in the different form factors \cite{Patrignani:2016xqp}.
To evaluate $t_+$, the values $m_D=1.870\:{\rm GeV}$ and $m_\pi=135\:{\rm MeV}$ should be used.}
\end{table}

\begin{table}
\begin{tabular}{lllll}
\hline\hline
            & & \hspace{1ex} Nominal fit  &  & Higher-order fit \\
\hline \\[-2.5ex]
$a_0^{f_\perp}$ & & $\wm 1.36\pm 0.07$ & & $\wm 1.32\pm 0.09$ \\ 
$a_1^{f_\perp}$ & & $-1.70\pm 0.83$ & & $-1.33\pm 0.98$ \\ 
$a_2^{f_\perp}$ & & $\wm 0.71\pm 4.34$ & & $-1.38\pm 8.60$ \\ 
$a_3^{f_\perp}$ & &  & & $\wm 7.02\pm 29.2$ \\
\hline \\[-2.5ex]
$a_0^{f_+}$ & & $\wm 0.83\pm 0.04$ & & $\wm 0.80\pm 0.05$ \\ 
$a_1^{f_+}$ & & $-2.33\pm 0.56$ & & $-1.94\pm 0.83$ \\ 
$a_2^{f_+}$ & & $\wm 8.41\pm 3.05$ & & $\wm 5.33\pm 8.04$ \\ 
$a_3^{f_+}$ & &  & & $\wm 10.1\pm 28.8$ \\ 
\hline \\[-2.5ex]
$a_0^{f_0}$ & & $\wm 0.84\pm 0.04$ & & $\wm 0.82\pm 0.05$ \\ 
$a_1^{f_0}$ & & $-2.57\pm 0.60$ & & $-2.42\pm 0.88$ \\ 
$a_2^{f_0}$ & & $\wm 9.87\pm 3.15$ & & $\wm 7.71\pm 8.19$ \\ 
$a_3^{f_0}$ & &  & & $\wm 9.30\pm 28.8$ \\ 
\hline \\[-2.5ex]
$a_0^{g_\perp,g_+}$ & & $\wm 0.69\pm 0.02$ & & $\wm 0.68\pm 0.03$ \\ 
$a_1^{g_\perp}$ & & $-0.68\pm 0.32$ & & $-0.89\pm 0.58$ \\ 
$a_2^{g_\perp}$ & & $\wm 0.70\pm 2.18$ & & $\wm 3.97\pm 6.81$ \\ 
$a_3^{g_\perp}$ & &  & & $-10.8\pm 25.2$ \\ 
\hline \\[-2.5ex]
$a_1^{g_+}$ & & $-0.90\pm 0.29$ & & $-1.07\pm 0.55$ \\ 
$a_2^{g_+}$ & & $\wm 2.25\pm 1.90$ & & $\wm 3.46\pm 6.42$ \\ 
$a_3^{g_+}$ & &  & & $\wm 0.49\pm 24.1$ \\ 
\hline \\[-2.5ex]
$a_0^{g_0}$ & & $\wm 0.73\pm 0.04$ & & $\wm 0.71\pm 0.05$ \\ 
$a_1^{g_0}$ & & $-0.97\pm 0.52$ & & $-0.93\pm 0.77$ \\ 
$a_2^{g_0}$ & & $\wm 0.83\pm 2.61$ & & $\wm 1.64\pm 7.87$ \\ 
$a_3^{g_0}$ & &  & & $-1.73\pm 28.0$ \\ 
\hline \\[-2.5ex]
$a_0^{h_\perp}$ & & $\wm 0.63\pm 0.03$ & & $\wm 0.62\pm 0.05$ \\ 
$a_1^{h_\perp}$ & & $-1.04\pm 0.45$ & & $-0.88\pm 0.72$ \\ 
$a_2^{h_\perp}$ & & $\wm 1.42\pm 2.67$ & & $\wm 1.42\pm 7.74$ \\ 
$a_3^{h_\perp}$ & &  & & $-0.41\pm 27.8$ \\ 
\hline \\[-2.5ex]
$a_0^{h_+}$ & & $\wm 1.11\pm 0.07$ & & $\wm 1.10\pm 0.10$ \\ 
$a_1^{h_+}$ & & $-0.69\pm 0.92$ & & $-0.56\pm 1.07$ \\ 
$a_2^{h_+}$ & & $-2.84\pm 5.19$ & & $-3.85\pm 9.28$ \\ 
$a_3^{h_+}$ & &  & & $\wm 5.61\pm 29.5$ \\ 
\hline \\[-2.5ex]
$a_0^{\widetilde{h}_\perp,\widetilde{h}_+}$ & & $\wm 0.63\pm 0.03$ & & $\wm 0.63\pm 0.05$ \\ 
$a_1^{\widetilde{h}_\perp}$ & & $-1.39\pm 0.58$ & & $-1.55\pm 0.81$ \\ 
$a_2^{\widetilde{h}_\perp}$ & & $\wm 4.22\pm 3.97$ & & $\wm 6.20\pm 8.12$ \\ 
$a_3^{\widetilde{h}_\perp}$ & &  & & $-5.19\pm 28.3$ \\ 
\hline \\[-2.5ex]
$a_1^{\widetilde{h}_+}$ & & $-1.19\pm 0.56$ & & $-1.23\pm 0.80$ \\ 
$a_2^{\widetilde{h}_+}$ & & $\wm 3.73\pm 3.73$ & & $\wm 4.36\pm 8.08$ \\ 
$a_3^{\widetilde{h}_+}$ & &  & & $-0.84\pm 28.1$ \\
\hline\hline
\end{tabular}
\caption{\label{tab:fitresults}Results for the $z$-expansion parameters describing the form factors in the physical limit via Eq.~(\protect\ref{eq:zexp}).
As explained in the main text, the nominal fit is used to evaluate the central values and statistical uncertainties, while the higher-order fit is used
to estimate systematic uncertainties. Because of the endpoint constraints (\protect\ref{eq:FFC3}) and (\protect\ref{eq:FFC4}) at $z=0$,
the parameters $a_0^{g_\perp,g_+}$ and $a_0^{\widetilde{h}_\perp,\widetilde{h}_+}$ are common to the form factors
$g_\perp$, $g_+$ and $\widetilde{h}_\perp$, $\widetilde{h}_+$, respectively.
Files containing the parameter values with more digits and the full covariance matrices are provided as supplemental material.}
\end{table}

\FloatBarrier
\section{\label{sec:Lcn}Phenomenology of \texorpdfstring{$\bm{\Lambda_c \to n\, \ell^+ \nu_\ell}$}{Lambda_c to n l+ nu}}
\FloatBarrier

\begin{figure}
 \includegraphics[width=0.49\linewidth]{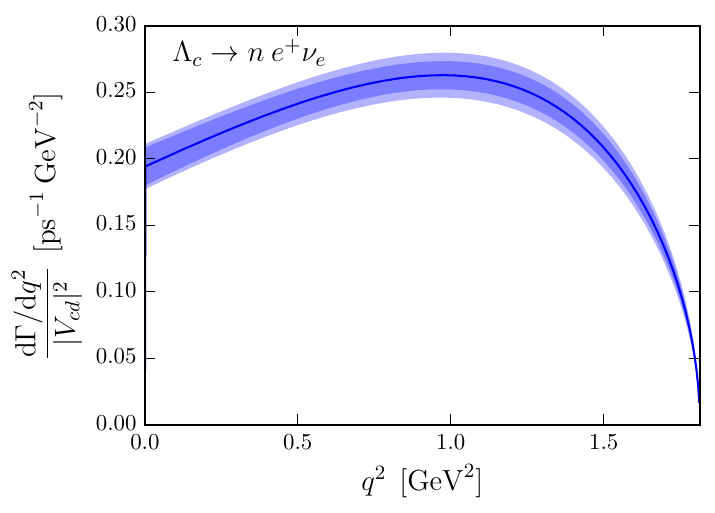} \hfill  \includegraphics[width=0.49\linewidth]{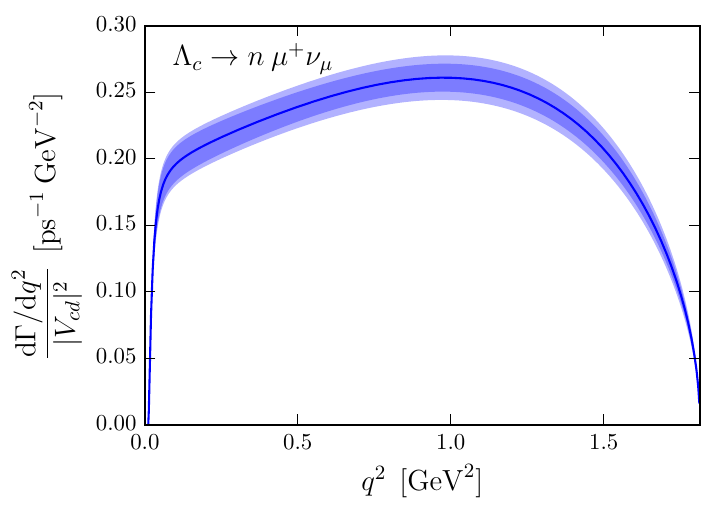}
 \caption{\label{fig:Lcn}Predictions for the $\Lambda_c \to n\, e^+ \nu_e$ and $\Lambda_c \to n\, \mu^+ \nu_\mu$
 differential decay rates in the Standard Model, without the factor of $|V_{cd}|^2$.
 The inner and outer bands show the statistical and total uncertainty originating from the form factors.}
\end{figure}

In terms of the helicity form factors, the $\Lambda_c \to n\, \ell^+ \nu_\ell$ differential decay rate in the Standard
Model reads
\begin{eqnarray}
\nonumber \frac{\mathrm{d}\Gamma}{\mathrm{d} q^2} &=&
\frac{G_F^2 |V_{cd}|^2 \sqrt{s_+s_-}
    }{768 \pi ^3 m_{\Lambda_c}^3 } \left(1-\frac{m_\ell^2}{q^2}\right)^2 \\
\nonumber     &&\times\Bigg\{4 \left(m_\ell^2+2 q^2\right) \left(
     s_+ \left[  g_\perp  \right]^2 + s_- \left[f_\perp\right]^2 \right)  \\
\nonumber    && \hspace{2ex} +2 \frac{m_\ell^2+2 q^2}{q^2} \left(s_+
   \left[\left(m_{\Lambda_c}-m_N\right) g_+ \right]^2+s_- \left[\left(m_{\Lambda_c}+m_N\right)f_+ \right]^2\right)\\
   &&\hspace{2ex} +\frac{6 m_\ell^2}{q^2} \left(s_+ \left[ \left(m_{\Lambda_c}-m_N\right) f_0
   \right]^2 + s_-
   \left[ \left(m_{\Lambda_c}+m_N\right) g_0  \right]^2\right)\Bigg\}.
\end{eqnarray}
Evaluating this expression for $\ell=e$ and $\ell =\mu$ using the form factor results described in the previous section gives
the results shown in Fig.~\ref{fig:Lcn}. The $q^2$-integrated rates are
\begin{eqnarray}
 \frac{\Gamma(\Lambda_c\to n\, e^+\nu_e)}{|V_{cd}|^2}
 &=&  \left( 0.405 \pm 0.016_{\,\rm stat} \pm 0.020_{\,\rm syst} \right) \:{\rm ps}^{-1}, \label{eq:GammaLcnenu}\\
 \frac{\Gamma(\Lambda_c\to n\, \mu^+\nu_\mu)}{|V_{cd}|^2}
 &=&  \left( 0.396 \pm 0.016_{\,\rm stat} \pm 0.020_{\,\rm syst} \right) \:{\rm ps}^{-1}, \label{eq:GammaLcnmunu}
\end{eqnarray}
with statistical and systematic uncertainties from the form factors. Using $|V_{cd}|=0.22497(67)$ from UTFit \cite{UTfit}
and the $\Lambda_c$ lifetime $\tau_{\Lambda_c}=0.200(6)\:{\rm ps}$ from the Particle Data Group \cite{Patrignani:2016xqp} yields
the branching fractions
\begin{eqnarray}
 \mathcal{B}(\Lambda_c\to n\, e^+\nu_e)
 &=& \left( 0.410 \pm 0.026_{\,\rm LQCD} \pm 0.012_{\,\tau_{\Lambda_c}} \pm 0.002_{\, |V_{cd}|} \right)\%, \\
 \mathcal{B}(\Lambda_c\to n\, \mu^+\nu_\mu)
 &=& \left( 0.400 \pm 0.026_{\,\rm LQCD} \pm 0.012_{\,\tau_{\Lambda_c}} \pm 0.002_{\, |V_{cd}|} \right)\%,
\end{eqnarray}
where the uncertainties labeled ``LQCD'' are the total uncertainties resulting from the form factors.

Previous predictions for $\Gamma(\Lambda_c\to n\, e^+\nu_e)/|V_{cd}|^2$ and $\Gamma(\Lambda_c\to n\, \mu^+\nu_\mu)/|V_{cd}|^2$
are summarized in Table \ref{tab:Lcn}. All references included there
predicted decay rates somewhat lower\footnote{Reference \cite{Azizi:2009wn} reported
values for $\Gamma/|V_{cd}|^2$ that are higher than calculated here by an order of magnitude, but these values appear to
be inconsistent with the form factor parameterizations given in the same work, and are therefore not included here.}
than the lattice QCD results (\ref{eq:GammaLcnenu}) and (\ref{eq:GammaLcnmunu}). While
Refs.~\cite{Ivanov:1996fj, Pervin:2005ve, Gutsche:2014zna, Faustov:2016yza} and \cite{Li:2016qai} estimated
the $\Lambda_c\to n$ form factors using quark models and light-cone sum rules, respectively, Ref.~\cite{Lu:2016ogy} derived
$\mathcal{B}(\Lambda_c\to n\, e^+\nu_e)$ from the BESIII experimental result for $\mathcal{B}(\Lambda_c\to\Lambda e^+\nu_e)$ \cite{Ablikim:2015prg}
using flavor $SU(3)$ symmetry. The low rate obtained in that way can likely be attributed in part to the assumption $m_n = m_\Lambda$,
which results in an underestimated phase space.

The authors of Ref.~\cite{Khodjamirian:2011jp} also calculated the $\Lambda_c \to N$ form factors using QCD light-cone sum rules,
but only at $q^2=0$. As shown in Table \ref{tab:LcNFF}, the results are consistent with the lattice QCD determination
within 1 to 2 $\sigma$.

\begin{table}
\begin{tabular}{lcccccc}
\hline\hline
Reference & \hspace{2ex} & Method & \hspace{2ex} & $\Gamma(\Lambda_c\to n\, e^+\nu_e)/|V_{cd}|^2$ [${\rm ps}^{-1}$] & \hspace{2ex} &  $\Gamma(\Lambda_c\to n\, \mu^+\nu_\mu)/|V_{cd}|^2$ [${\rm ps}^{-1}$]   \\
\hline
Ivanov {\it et al.}, 1996 \cite{Ivanov:1996fj}             &&   Quark model           &&   $0.26$              &&                    \\
Pervin {\it et al.}, 2005 \cite{Pervin:2005ve}             &&   Quark model           &&   $0.203$, $0.269$    &&                    \\
Gutsche {\it et al.}, 2014 \cite{Gutsche:2014zna}          &&   Quark model           &&   $0.20$              &&  $0.19$            \\
L\"u {\it et al.}, 2016 \cite{Lu:2016ogy}                  &&   $SU(3)$ symmetry      &&   $0.289 \pm 0.035$   &&                    \\
Faustov and Galkin, 2016 \cite{Faustov:2016yza}            &&   Quark model           &&   $0.265$             &&  $0.260$           \\
Li {\it et al.}, 2016 \cite{Li:2016qai}                    &&   Light-cone sum rules  &&   $0.267 \pm 0.011$   &&                    \\[1ex]
This work                                                  &&   Lattice QCD           &&   $0.405 \pm 0.026$   &&  $0.396 \pm 0.025$ \\
\hline\hline
\end{tabular}
\caption{\label{tab:Lcn}Comparison of predictions for $\Gamma(\Lambda_c\to n\, e^+\nu_e)/|V_{cd}|^2$ and $\Gamma(\Lambda_c\to n\, \mu^+\nu_\mu)/|V_{cd}|^2$
with the lattice QCD results obtained here. The two different values from Ref.~\cite{Pervin:2005ve} correspond to nonrelativistic and
semirelativistic kinetic terms in the model. Reference \cite{Lu:2016ogy} predicted the branching fraction, which was converted here
using $|V_{cd}|=0.22497(67)$ \cite{UTfit} and $\tau_{\Lambda_c}=0.200(6)\:{\rm ps}$ \cite{Patrignani:2016xqp}. Reference \cite{Li:2016qai}
gave a value for $\Gamma$, which was converted using the value of $|V_{cd}|$ given there.}
\end{table}

\begin{table}
\begin{tabular}{lccccccccccc}
\hline\hline
Reference & \hspace{1ex} & Method   & \hspace{1ex} &  $f_1(0)$ & \hspace{1ex} &  $f_2(0)$ & \hspace{1ex} &  $g_1(0)$ & \hspace{1ex} &  $g_2(0)$   \\
\hline
\multirow{2}{*}{A.~Khodjamirian {\it et al.}, 2011 \cite{Khodjamirian:2011jp}}
&& LCSR, $\eta_{\Lambda_c}^{(\mathcal{A})}$   &&   $0.46^{+0.15}_{-0.11}$  && $-0.32^{+0.08}_{-0.07}$ && $0.49^{+0.14}_{-0.11}$ && $-0.20^{+0.09}_{-0.06}$  \\[1ex]
&& LCSR, $\eta_{\Lambda_c}^{(\mathcal{P})}$   &&   $0.59^{+0.15}_{-0.16}$  && $-0.43^{+0.13}_{-0.12}$ && $0.55^{+0.14}_{-0.15}$ && $-0.16^{+0.08}_{-0.05}$  \\[1ex]
This work             &&   Lattice QCD        &&   $0.672 \pm 0.039$       && $-0.321 \pm 0.038$      && $0.602 \pm  0.031$     && $0.003 \pm 0.052$     \\
\hline\hline
\end{tabular}
\caption{\label{tab:LcNFF} Comparison of $\Lambda_c \to N$ form factor values at $q^2=0$ calculated in Ref.~\cite{Khodjamirian:2011jp} using
QCD light-cone sum rules for two different $\Lambda_c$ interpolating fields, $\eta_{\Lambda_c}^{(\mathcal{A})}$ and $\eta_{\Lambda_c}^{(\mathcal{P})}$,
with the present lattice QCD results. The form factors defined in Ref.~\cite{Khodjamirian:2011jp} are related to the helicity form factors
as $f_1=[ (m_{\Lambda_c}+m_N)^2 f_+  -  q^2 f_\perp]/s_+$, $f_2=m_{\Lambda_c} (m_{\Lambda_c}+m_N)(f_+ - f_\perp)/s_+$,
$g_1=[(m_{\Lambda_c}-m_N)^2 g_+ - q^2 g_\perp]/s_-$, and $g_2=m_{\Lambda_c} (m_{\Lambda_c}-m_N)(g_+-g_\perp)/s_-$.}
\end{table}

\FloatBarrier
\section{\label{sec:Lcp}Phenomenology of \texorpdfstring{$\bm{\Lambda_c \to p\,\mu^+\mu^-}$}{Lambda_c to p mu+ mu-}}
\FloatBarrier

The theory of exclusive $c \to u \ell^+ \ell^-$ transitions in the Standard Model is complicated by the dominance of nonlocal hadronic
matrix elements that cannot easily be calculated in lattice QCD. The form factors
computed in this work only describe local matrix elements of the form $\langle p | \bar{u}\Gamma c|\Lambda_c\rangle$.
In the following, two different approaches for expressing the $\Lambda_c \to p\,\mu^+\mu^-$ observables in terms of these form factors
will be considered: a perturbative calculation of effective Wilson coefficients at next-to-next-to-leading order (Sec.~\ref{sec:LcpPT}),
and a phenomenological Breit-Wigner model for the contributions of intermediate $\phi$, $\omega$, and $\rho^0$ resonances (Sec.~\ref{sec:LCpBW}).
The resulting predictions for the $\Lambda_c \to p\,\mu^+\mu^-$ differential branching fraction and angular distribution are given in
Sec.~\ref{sec:LCpObs}. In the effective-field-theory description, possible heavy new physics beyond the Standard Model contributes
only via the local matrix elements given by the form factors. In the case of $\Lambda_c \to p\,\mu^+\mu^-$ (but not, for example,
in lepton-flavor-violating modes such as $\Lambda_c \to p\,e^+ \mu^-$), such contributions still interfere with 
Standard-Model contributions. Nevertheless, the $\Lambda_c \to p\,\mu^+\mu^-$ forward-backward asymmetry is nonzero only in the presence of new physics,
and therefore provides a clean test of the Standard Model.

\FloatBarrier
\subsection{\label{sec:LcpPT}Standard-Model Wilson coefficients in perturbation theory}
\FloatBarrier

The $c\to u \ell^+\ell^-$ effective weak Lagrangian at $\mu < m_b$, after integrating out the $b$ quark, has the form
\begin{equation}
\mathcal{L}_{\rm eff} = \frac{4 G_F}{\sqrt 2}\sum_{q=d,s}V_{cq}^*V_{uq}\left(\tilde{C}_1 P_1^{(q)} +\tilde{C}_2 P_2^{(q)}  + \sum_{i=3}^{10}\tilde{C}_i P_i \right),
\end{equation}
where $P_{1,...,6}$ are four-quark operators, $P_7$ and $P_8$ are electromagnetic and gluonic dipole operators,
and $P_9$, $P_{10}$ are semileptonic operators  \cite{deBoer:2015boa, deBoer:2016dcg}.
Following Refs.~\cite{deBoer:2015boa, deBoerphdthesis}, the matrix elements are written as
\begin{equation}
 \langle \mathcal{L}_{\rm eff} \rangle = \frac{4\,G_F}{\sqrt{2}} \frac{\alpha_e}{4\pi} \sum_i C_i(q^2)\, \langle Q_i \rangle, \label{eq:LeffPheno}
\end{equation}
with rescaled electromagnetic dipole and semileptonic operators
\begin{eqnarray}
 Q_7&=&\frac{m_c}{e}(\bar u\sigma^{\mu\nu}P_Rc)F_{\mu\nu},\label{eq:Q7}\\
 Q_9&=&(\bar u\gamma_{\mu}P_Lc)\left(\overline \ell\gamma^{\mu}\ell\right), \label{eq:Q9}\\
 Q_{10}&=&(\bar u\gamma_{\mu}P_Lc)\left(\overline \ell \gamma^{\mu}\gamma_5 \ell\right), \label{eq:Q10}
\end{eqnarray}
and $q^2$-dependent effective Wilson coefficients
\begin{eqnarray}
C_{7,9}(q^2)&=&\frac{4\pi}{\alpha_s }\left[ V_{cd}^*V_{ud}\, C_{7,9}^{\text{eff}(d)}(q^2)+ V_{cs}^*V_{us}\, C_{7,9}^{\text{eff}(s)}(q^2) \right],
\end{eqnarray}
which include the perturbative matrix elements of the four-quark and gluonic dipole operators. The Wilson coefficient $C_{10}$ is zero in the Standard Model
due to CKM unitarity, since all down-type quarks are treated as massless at $\mu=m_W$ and $C_{10}$ does not mix under renormalization
\cite{deBoer:2016dcg, deBoerphdthesis}. At next-to-next-to-leading order, including the recently derived two-loop QCD matrix elements
of $P_1^{(q)}$ and $P_2^{(q)}$ for arbitrary momentum transfer \cite{deBoer:2017way, deBoerphdthesis}, the effective Wilson coefficients are given by
\begin{eqnarray}
 C_9^{\text{eff}(q)}(q^2)&=&\tilde{C}_9^{(0+1+2)}+\frac{\alpha_s}{4\pi}\Bigg[ \frac{8}{27}\tilde{C}_1+\frac{2}{9}\tilde{C}_2-\frac{8}{9}\tilde{C}_3-\frac{32}{27}\tilde{C}_4-\frac{128}{9}\tilde{C}_5-\frac{512}{27}\tilde{C}_6\nonumber\\
 &&+L(m_c^2,q^2)\left(\frac{28}{9}\tilde{C}_3+\frac{16}{27}\tilde{C}_4+\frac{304}{9}\tilde{C}_5+\frac{256}{27}\tilde{C}_6\right)\nonumber\\
 &&+L(m_s^2,q^2)\left(-\frac{4}{3}\tilde{C}_3-\frac{40}{3}\tilde{C}_5\right)\nonumber\\
 &&+L(0,q^2)\left(\frac{16}{9}\tilde{C}_3+\frac{16}{27}\tilde{C}_4+\frac{184}{9}\tilde{C}_5+\frac{256}{27}\tilde{C}_6\right)\nonumber\\
 &&+\left(\delta_{qs}L(m_s^2,q^2)+\delta_{qd}L(0,q^2)\right)\left(-\frac{8}{27}\tilde{C}_1-\frac{2}{9}\tilde{C}_2\right)\Bigg]^{(0+1)} \nonumber\\
 &&+\left(\frac{\alpha_s}{4\pi}\right)^2\Bigg[ F_{1,q}^{(9)}(m_c^2, q^2)\,\tilde{C}_1^{(0)} + F_{2,q}^{(9)}(m_c^2, q^2)\,\tilde{C}_2^{(0)} + F_8^{(9)}(m_c^2, q^2)\, C_8^\text{eff}  \Bigg] \label{eq:C9_eff}
\end{eqnarray}
and
\begin{eqnarray}
 C_7^{\text{eff}(q)}(q^2)&=&\tilde{C}_7^{(0+1+2)}+\frac{\alpha_s}{4\pi}\Bigg[\frac{2}{3}\tilde{C}_3 + \frac{8}{9}\tilde{C}_4 + \frac{40}{3}\tilde{C}_5 + \frac{160}{9} \tilde{C}_6 \Bigg]^{(0+1)} \nonumber\\
 &&+\left(\frac{\alpha_s}{4\pi}\right)^2   \Bigg[\left(-\frac{1}{6}\tilde{C}_1^{(0)}+\tilde{C}_2^{(0)}\right)\,F_{2,q}^{(7)}(m_c^2, q^2)+F_8^{(7)}(m_c^2,q^2) \, C_8^\text{eff}\Bigg], \label{eq:C7_eff}
\end{eqnarray}
where the Wilson coefficients $\tilde{C}_i$ are expanded in the strong coupling as
\begin{equation}
  \tilde{C}_i=\tilde{C}_i^{(0)}+\frac{\alpha_s}{4\pi}\tilde{C}_i^{(1)}+\left(\frac{\alpha_s}{4\pi}\right)^2\tilde{C}_i^{(2)}+...\, ,
\end{equation}
and the notation $\tilde{C}_i^{(0+1)} \equiv \tilde{C}_i^{(0)}+\frac{\alpha_s}{4\pi}\tilde{C}_i^{(1)}$ etc. is used. Above,
\begin{equation}
 C_8^\text{eff}=\tilde{C}_8^{(1)}+ \tilde{C}_3^{(0)} -\frac{1}{6}\tilde{C}_4^{(0)} + 20\, \tilde{C}_5^{(0)} -\frac{10}{3} \tilde{C}_6^{(0)}.
\end{equation}
The functions $L(m^2,q^2)$, $L(0,q^2)$, $F_8^{(7)}(m_c^2, q^2)$, and $F_8^{(9)}(m_c^2, q^2)$ can be found in Appendix B of Ref.~\cite{deBoer:2015boa}.
The functions $F_{1,q}^{(9)}(m_c^2, q^2)$, $F_{2,q}^{(9)}(m_c^2, q^2)$, and $F_{1,q}^{(7)}(m_c^2, q^2) = -\frac{1}{6} F_{2,q}^{(7)}(m_c^2, q^2)$ are given
in Ref.~\cite{deBoer:2017way}, and were evaluated here using the files \texttt{fit\_F*.dat} provided as supplemental material in Ref.~\cite{deBoer:2017way}.
The values of $\tilde{C}_i^{(0)}$, $\frac{\alpha_s}{4\pi}\tilde{C}_i^{(1)}$, and $\left(\frac{\alpha_s}{4\pi}\right)^2\tilde{C}_i^{(2)}$
at $\mu = m_c$ were taken from Table 2.2 of Ref.~\cite{deBoerphdthesis}. For the purpose of estimating the perturbative uncertainties in the observables,
the values of these coefficients were provided by Stefan de Boer additionally for $\mu = \sqrt{2}\, m_c$ and $\mu = m_c/\sqrt{2}$ \cite{deBoerWC}.
The low-energy value of $\alpha_e$ was used, and the strong coupling $\alpha_s(\mu)$ for four flavors was evaluated using the \texttt{RunDec} package
\cite{Chetyrkin:2000yt}. The quark masses were set to $m_c^{\overline{\rm MS}}=1.27$ GeV, $m_c^{\rm pole}=1.47$ GeV,
$m_s^{\rm pole}=0.13$ GeV as in Ref.~\cite{deBoer:2017way}. The CKM matrix elements were taken from UTFit \cite{UTfit}.

\FloatBarrier
\subsection{\label{sec:LCpBW}Breit-Wigner model of resonant contributions}
\FloatBarrier

Similarly to Ref.~\cite{deBoer:2015boa}, the contributions from intermediate $\phi$, $\omega$, and $\rho^0$ resonances were modeled
using an effective Wilson coefficient $C_9^{\rm R}(q^2)$ given by
\begin{equation}
C_9^{\rm R}(q^2) = a_\omega\, e^{i\delta_\omega} \left(\frac{1}{q^2-m_\omega^2+i m_\omega\Gamma_\omega}-\frac{3}{q^2-m_\rho^2+i m_\rho\Gamma_\rho}\right)+\frac{a_\phi\, e^{i\delta_\phi}}{q^2-m_\phi^2+im_\phi\Gamma_\phi}, \label{eq:C9R}
\end{equation}
and setting $C_7(q^2)=0$. Here, the relative magnitude and sign between the $\rho^0$ and $\omega$ amplitudes were fixed as for $D^+ \to \pi^+ \ell^+\ell^-$ in Ref.~\cite{Fajfer:2005ke}.
In the case of $\Lambda_c \to p\, \ell^+\ell^-$, the quark flow diagrams are different, but those diagrams that are expected to dominate
yield the same relation. The resonance masses and widths were taken from the Particle Data Group \cite{Patrignani:2016xqp}. To determine the couplings
$a_\omega$ and $a_\phi$, the $\Lambda_c \to p\,\mu^+\mu^-$ branching fraction was computed using only $C_9^{\rm R}(q^2)$ and keeping only the $\omega$
or $\phi$ contribution, and demanding that
\begin{equation}
\mathcal{B}(\Lambda_c (\to p V) \to p\,\mu^+\mu^-) \:=\: \mathcal{B}(\Lambda_c \to p\,V)\,\mathcal{B}(V\to \mu^+\mu^-) \hspace{4ex}\text{for}\hspace{2ex} V=\omega,\phi,
\end{equation}
where the right-hand side was evaluated using the following experimental inputs:
\begin{eqnarray}
 \mathcal{B}(\Lambda_c \to p\,\phi)   &=& (1.08 \pm 0.14) \times 10^{-3} \hspace{2ex}\text{\cite{Patrignani:2016xqp}}, \\
 \mathcal{B}(\phi \to \mu^+\mu^-)     &=& (2.87 \pm 0.19) \times 10^{-4} \hspace{2ex}\text{\cite{Patrignani:2016xqp}}, \\
 \frac{\mathcal{B}(\Lambda_c \to p\,\omega)\mathcal{B}(\omega \to \mu^+\mu^-) }{\mathcal{B}(\Lambda_c \to p\,\phi)\mathcal{B}(\phi \to \mu^+\mu^-)} &=& 0.23 \pm 0.08 \pm 0.03  \hspace{2ex}\text{\cite{Aaij:2017nsd}}.
\end{eqnarray}
This procedure gives
\begin{eqnarray}
 a_\omega &=& 0.068 \pm 0.013,  \label{eq:aomega} \\
 a_\phi   &=& 0.111 \pm 0.008.  \label{eq:aphi}
\end{eqnarray}
The phases $\delta_\omega$ and $\delta_\phi$ were varied independently in the ranges $0$ to $2\pi$ when calculating the
$\Lambda_c \to p\,\mu^+\mu^-$ observables presented in the next section.

\FloatBarrier
\subsection{\label{sec:LCpObs}Results for the \texorpdfstring{$\bm{\Lambda_c \to p\,\mu^+\mu^-}$}{Lambda_c to p mu+ mu-} observables}
\FloatBarrier

The two-fold differential decay rate of $\Lambda_c \to p\,\ell^+\ell^-$ with unpolarized $\Lambda_c$ can be written as
\begin{equation}
\frac{\mathrm{d}^2 \Gamma}{\mathrm{d} q^2\: \mathrm{d} \cos\theta_\ell}
 =          \frac{3}{2} \left( K_{1ss} \sin^2\theta_\ell +\, K_{1cc} \cos^2\theta_\ell + K_{1c} \cos\theta_\ell \right),
\end{equation}
where $\theta_\ell$ is the angle of the $\ell^+$ in the dilepton rest frame with respect to the direction of flight of the dilepton system
in the $\Lambda_c$ rest frame, and the coefficients $K_{1ss}$, $K_{1cc}$, and $K_{1c}$ are functions of $q^2$. The $q^2$-differential decay rate
is obtained by integrating over $\cos \theta_\ell$,
\begin{equation}
 \frac{\mathrm{d}\Gamma}{\mathrm{d}q^2}=2 K_{1ss} + K_{1cc}.
\end{equation}
The fraction of longitudinally polarized dimuons and the forward-backward asymmetry are defined as
\begin{equation}
 F_L = \frac{2 K_{1ss}- K_{1cc}}{\mathrm{d}\Gamma/\mathrm{d}q^2}
\end{equation}
and
\begin{equation}
 A_{FB} = \frac32 \frac{K_{1c}}{\mathrm{d}\Gamma/\mathrm{d}q^2}.
\end{equation}
For the effective Lagrangian (\ref{eq:LeffPheno}) with operators $Q_7$, $Q_9$, $Q_{10}$, the expressions for
$K_{1ss}$, $K_{1cc}$, and $K_{1c}$ in terms of the form factors and the Wilson coefficients $C_i(q^2)$ can be obtained from
Ref.~\cite{Boer:2014kda} (in the approximation $m_\ell=0$) or Ref.~\cite{Gutsche:2013pp} (for $m_\ell\neq 0$; used here). These references consider
the similar process $\Lambda_b \to \Lambda \ell^+\ell^-$; to adapt the equations to $\Lambda_c \to p\,\ell^+\ell^-$,
the factor of $|V_{tb}V_{ts}^*|^2$ needs to be removed and the masses need to replaced appropriately.

\begin{figure}
 \includegraphics[width=0.485\linewidth]{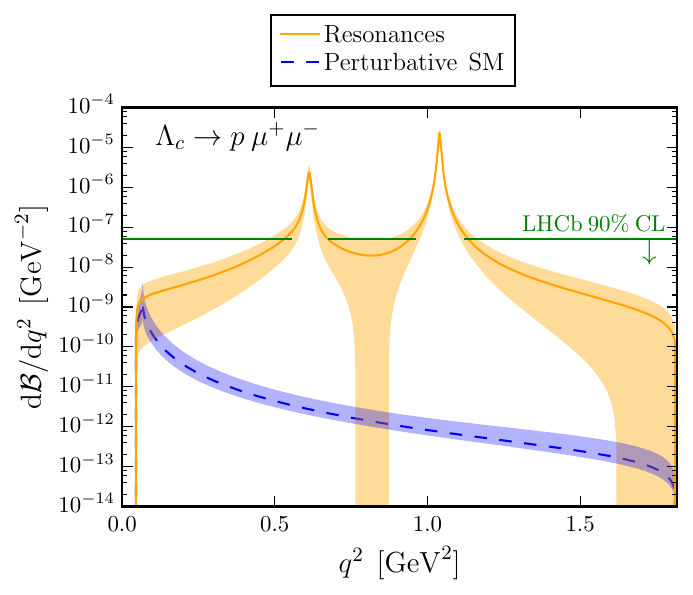} \hfill  \includegraphics[width=0.485\linewidth]{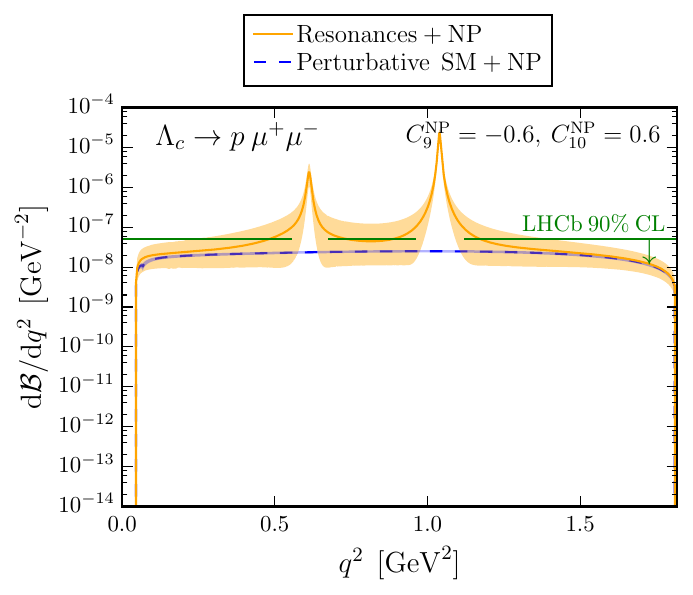}
 
\hspace{2ex} \includegraphics[width=0.46\linewidth]{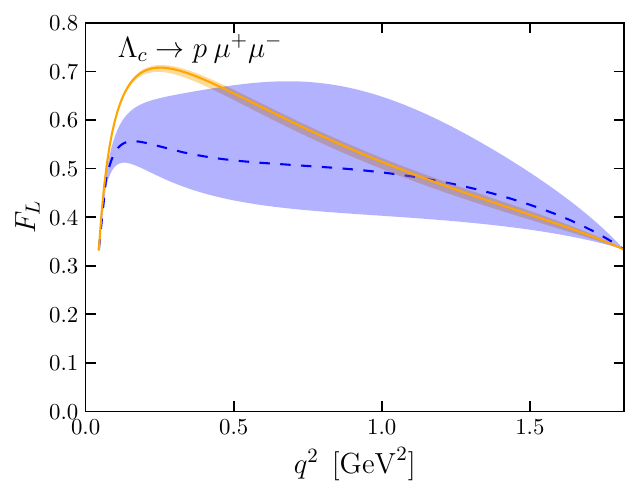} \hfill  \includegraphics[width=0.46\linewidth]{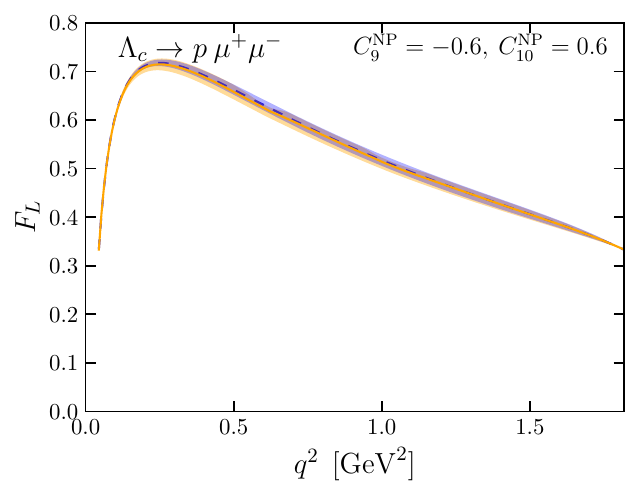}
 
\hspace{2ex} \includegraphics[width=0.46\linewidth]{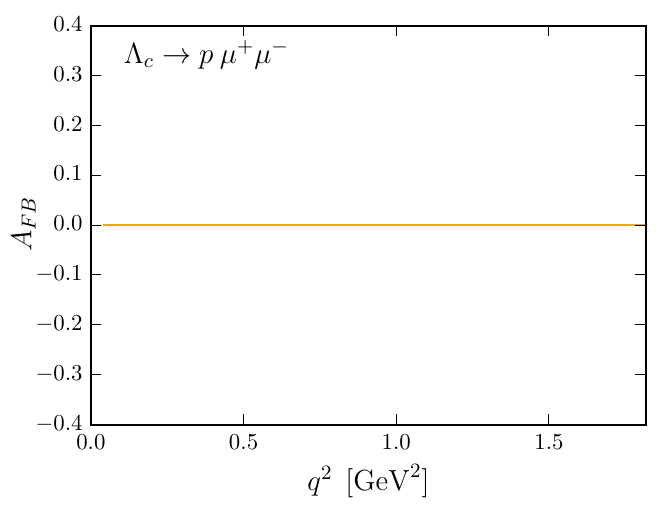}  \hfill \includegraphics[width=0.46\linewidth]{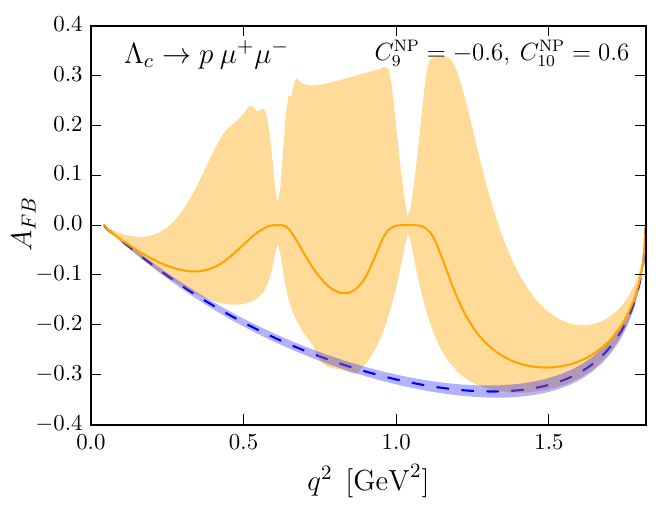}

\caption{\label{fig:Lcp}Predictions for the $\Lambda_c \to p\,\mu^+\mu^-$ differential branching fraction, fraction of longitudinally polarized dimuons, and
forward-backward asymmetry, in the Standard Model (left panels) and with a new-physics contribution of $C_9^{\rm NP}=-0.6,\: C_{10}^{\rm NP}=0.6$ (right panels).
Also indicated is the LHCb upper limit (at 90\% confidence level) on $\mathcal{B}(\Lambda_c \to p\,\mu^+\mu^-)$ in the $\sqrt{q^2}$ region excluding $\pm40$ MeV
intervals around $m_\omega$ and $m_\phi$ \cite{Aaij:2017nsd}.
The blue dashed curves use the perturbative results (\protect\ref{eq:C9_eff}) and (\protect\ref{eq:C7_eff}) for the Standard-Model contribution, with error
bands including the perturbative scale uncertainties and the form factor uncertainties. The orange solid curves instead use the resonant
Breit-Wigner model (\ref{eq:C9R}) for the Standard-Model contribution, with error bands including the changes in the observables under
independent variations of the phases $\delta_\omega$ and $\delta_\phi$ in the ranges $0$ to $2\pi$, as well as the uncertainties in the couplings
$a_\omega$ and $a_\phi$ and the form factor uncertainties. The curves show the averages over all values of phases.}
\end{figure}

The predictions for $\mathrm{d}\mathcal{B}/\mathrm{d}q^2 = \tau_{\Lambda_c}\mathrm{d}\Gamma/\mathrm{d}q^2$, $F_L$, and $A_{FB}$ for $\Lambda_c \to p\,\mu^+\mu^-$,
using either the perturbative Wilson coefficients (\protect\ref{eq:C9_eff}) and (\protect\ref{eq:C7_eff}) or the resonant model (\ref{eq:C9R}),
are shown in the left panels of Fig.~\ref{fig:Lcp}. In the right panels, an example new-physics contribution of
$C_9^{\rm NP}=-0.6,\: C_{10}^{\rm NP}=0.6$ was added to the Wilson coefficients to illustrate the effect on the observables.
A contribution of this magnitude is not yet excluded by experimental measurements of rare charm meson decays \cite{deBoer:2015boa}.
Also shown in Fig.~\ref{fig:Lcp} is the LHCb upper limit of $\mathcal{B}(\Lambda_c \to p\,\mu^+\mu^-) < 7.7\times 10^{-8}$ (at 90\% confidence level)
in the $\sqrt{q^2}$ region excluding $\pm40$ MeV intervals around $m_\omega$ and $m_\phi$ \cite{Aaij:2017nsd}.
The LHCb upper limit on $\mathcal{B}(\Lambda_c \to p\,\mu^+\mu^-)$ also does not yet exclude the new-physics scenario considered here,
but comes close.

The error bands of the nonresonant SM predictions are dominated by the perturbative uncertainty, which was estimated
by computing the changes in the observables when varying the renormalization scale from $\mu=m_c$ to $\mu=\sqrt{2}\, m_c$ and $\mu=m_c/\sqrt{2}$.
While doing this scale variation, the renormalization-group running of the tensor form factors was included for consistency, by multiplying
these form factors with
\begin{equation}
 \left(\frac{\alpha_s(\mu)}{\alpha_s(m_c)} \right)^{-\gamma_T^{(0)}/(2\beta_0)},
\end{equation}
where $\gamma_T^{(0)}=2\, C_F = 8/3$ is the anomalous dimension of the tensor current \cite{Broadhurst:1994se}
and $\beta_0=(11\, N_c - 2\, N_f)/3=25/3$ is the leading-order coefficient of the QCD beta function for 4 active flavors.
The error bands of the predictions using the resonant model (\ref{eq:C9R}) are dominated by the phase uncertainty,
which was estimated by independently varying $\delta_\omega$ and $\delta_\phi$ in the ranges $0$ to $2\pi$ and showing
the resulting ranges of the observables, and the uncertainties in the couplings $a_\omega$ and $a_\phi$ as given
in Eqs.~(\ref{eq:aomega}) and (\ref{eq:aphi}).

The branching fractions integrated over the entire $q^2$ range are found to be
\begin{eqnarray}
 \mathcal{B}(\Lambda_c \to p\,\mu^+\mu^-)_{\text{Perturbative SM}}&=& \left(4.1 \pm 0.4 ^{+6.1}_{-1.9} \right)\times10^{-11}, \label{eq:BLcp} \\
 \mathcal{B}(\Lambda_c \to p\,\mu^+\mu^-)_{\text{Resonances}}      &=& \left(3.7 ^{+1.1}_{-1.2} \right)\times10^{-7},
\end{eqnarray}
where, for the perturbative SM prediction, the first uncertainty given is the form factor uncertainty from the lattice
calculation, and the second uncertainty is the perturbative uncertainty estimated by varying the renormalization scale.
As can be seen in Fig.~\ref{fig:Lcp}, the low-$q^2$ region gives most of the perturbative SM contribution.

The value for $\mathcal{B}(\Lambda_c \to p\,\mu^+\mu^-)_{\text{Perturbative SM}}$ calculated here is approximately
$10^3$ times higher than that obtained in Refs.~\cite{Azizi:2010zzb, Sirvanli:2016wnr}. While \cite{Azizi:2010zzb}
does not give a reference for the Wilson coefficients, \cite{Sirvanli:2016wnr} reportedly uses Wilson
coefficients from Ref.~\cite{Paul:2011ar}. The Wilson coefficients from Ref.~\cite{Paul:2011ar} actually tend to
give higher branching ratios than the Wilson coefficients employed here \cite{deBoer:2016dcg, deBoer:2017way, deBoerphdthesis, deBoerWC},
and the very small branching fractions obtained in \cite{Azizi:2010zzb, Sirvanli:2016wnr} are puzzling. Note that
Refs.~\cite{Azizi:2010zzb, Sirvanli:2016wnr} write the matrix elements of the tensor current in terms of six form
factors, of which only four are independent. The numerical parameterizations given there for the six tensor form factors violate
the exact kinematical relations between these form factors.

The $\Lambda_c \to p\,\mu^+\mu^-$ forward-backward asymmetry, shown at the bottom of Fig.~\ref{fig:Lcp},
vanishes in the Standard Model because it contains an overall factor of $C_{10}$.
New physics giving a nonzero $C_{10}$ would produce a nonzero forward-backward asymmetry, as shown in the bottom-right panel of Fig.~\ref{fig:Lcp}.
While the actual values of $A_{FB}$ strongly depend on the details of the resonant contributions to $C_9$, this observable still provides a clean
null test of the Standard Model.

\FloatBarrier

\section{\label{sec:conclusions}Conclusions}

In this paper, a precise lattice QCD determination of the $\Lambda_c \to N$ ($N=n,p$) vector, axial vector, and tensor form factors
was reported. The results provide Standard-Model predictions for the $\Lambda_c \to n\, e^+ \nu_e$ and
$\Lambda_c \to n\, \mu^+ \nu_\mu$ decay rates with an uncertainty of 6.4\%. The rates calculated here for these decays
are higher than those predicted in Refs.~\cite{Ivanov:1996fj, Pervin:2005ve, Gutsche:2014zna, Lu:2016ogy, Faustov:2016yza, Li:2016qai}
using quark models, sum rules, or $SU(3)$ symmetry, by factors ranging from 1.4 to 2.

The form factors were then applied to study the differential branching fraction and angular distribution
of the rare charm decay $\Lambda_c \to p\,\mu^+\mu^-$, using either perturbative results for the
effective Wilson coefficients in the Standard Model \cite{deBoer:2016dcg, deBoer:2017way, deBoerphdthesis}
or a simple Breit-Wigner model for the long-distance contributions from the $\phi$, $\omega$, and $\rho^0$ resonances.
The perturbative analysis gives
$\mathcal{B}(\Lambda_c \to p\,\mu^+\mu^-)_{\text{Perturbative SM}}= \left(4.1 \pm 0.4 ^{+6.1}_{-1.9} \right)\times10^{-11}$.
The LHCb upper limit of $\mathcal{B}(\Lambda_c \to p\,\mu^+\mu^-) < 7.7\times 10^{-8}$ (at 90\% confidence level)
in the dimuon mass region excluding $\pm40$ MeV intervals around $m_\omega$ and $m_\phi$ \cite{Aaij:2017nsd} still
allows new-physics contributions to $C_9$ and $C_{10}$ [defined as in Eq.~(\ref{eq:LeffPheno})] of order $\mathcal{O}(1)$.

The $\Lambda_c \to p\,\mu^+\mu^-$ observables in the Standard Model are dominated by long-distance contributions from nonlocal matrix elements,
whose treatment is still very unsatisfactory. However, the forward-backward asymmetry is nonzero only in the presence
of new physics, and a measurement would provide a clean test of the Standard Model.
More detailed phenomenological studies, including other observables such as CP asymmetries and lepton-flavor-violating decay modes, are warranted.

\vspace{10ex}

\begin{acknowledgments}
  I thank Stefan de Boer for providing the values of the $c\to u\ell^+\ell^-$ Wilson coefficients for additional
  choices of the renormalization scale, and Christoph Lehner for computing the perturbative lattice-to-continuum matching and
  $\mathcal{O}(a)$-improvement coefficients for the $c\to u$ currents. I am grateful to the RBC and UKQCD Collaborations for making their
  gauge field ensembles available. This work was supported by National Science Foundation Grant No.~PHY-1520996 and by the RHIC Physics Fellow Program
  of the RIKEN BNL Research Center.  High-performance computing resources were provided by the Extreme Science and Engineering Discovery Environment (XSEDE),
  supported by National Science Foundation Grant No.~ACI-1053575, as well as the National Energy Research Scientific Computing Center, a DOE Office of Science
  User Facility supported by the Office of Science of the U.S. Department of Energy under Contract No. DE-AC02-05CH11231.
  The Chroma software system \cite{Edwards:2004sx} was used for the lattice calculations.
\end{acknowledgments}

\section*{Note added}

In this version, I corrected an error in the lepton-mass contributions to $F_L$ in Fig.~\ref{fig:Lcp}. I thank Gudrun Hiller for bringing this to my attention.

\newpage

\appendix

\FloatBarrier
\section*{\label{sec:FFlat}Appendix: Lattice form factor data}
\FloatBarrier

\begin{table}[!h]
\footnotesize
\begin{tabular}{ccccllllllllllllll}
\hline\hline
         & & $|\mathbf{p'}|^2/(2\pi/L)^2$ && \hspace{2.5ex} \texttt{C14} & \hspace{2ex} & \hspace{2.5ex} \texttt{C24} & \hspace{2ex} & \hspace{2.5ex} \texttt{C54} & \hspace{2ex} & \hspace{2.5ex} \texttt{F23} & \hspace{2ex} & \hspace{2.5ex} \texttt{F43} & \hspace{2ex} & \hspace{2.5ex} \texttt{F63} \\
\hline
$f_\perp$              && 1 &&  1.884(54) &&  1.897(35) &&  1.941(29) &&  1.869(43) &&  1.895(32) &&  1.947(33)  \\ 
                       && 2 &&  1.567(44) &&  1.587(31) &&  1.629(28) &&  1.574(37) &&  1.599(30) &&  1.655(28)  \\ 
                       && 3 &&  1.287(60) &&  1.300(47) &&  1.348(41) &&  1.290(49) &&  1.329(46) &&  1.396(40)  \\ 
                       && 4 &&  1.122(41) &&  1.153(21) &&  1.190(21) &&  1.171(28) &&  1.185(23) &&  1.223(26)  \\ 
                       && 5 &&  0.998(38) &&  1.025(27) &&  1.061(28) &&  1.055(23) &&  1.073(20) &&  1.113(21)  \\ 
\hline
$f_+$                  && 1 &&  1.093(22) &&  1.085(16) &&  1.087(14) &&  1.070(13) &&  1.074(12) &&  1.087(15)  \\ 
                       && 2 &&  0.883(22) &&  0.887(17) &&  0.894(15) &&  0.875(18) &&  0.884(15) &&  0.902(18)  \\ 
                       && 3 &&  0.720(24) &&  0.725(20) &&  0.739(17) &&  0.706(22) &&  0.726(18) &&  0.752(20)  \\ 
                       && 4 &&  0.638(25) &&  0.652(20) &&  0.658(17) &&  0.675(20) &&  0.667(18) &&  0.666(21)  \\ 
                       && 5 &&  0.580(41) &&  0.601(47) &&  0.605(48) &&  0.613(28) &&  0.613(32) &&  0.620(24)  \\ 
\hline
$f_0$                  && 1 &&  0.983(20) &&  0.969(17) &&  0.971(15) &&  0.967(16) &&  0.966(14) &&  0.970(17)  \\ 
                       && 2 &&  0.822(20) &&  0.828(17) &&  0.833(15) &&  0.812(18) &&  0.824(16) &&  0.839(19)  \\ 
                       && 3 &&  0.694(22) &&  0.701(19) &&  0.714(17) &&  0.680(22) &&  0.701(19) &&  0.726(19)  \\ 
                       && 4 &&  0.635(23) &&  0.650(19) &&  0.657(17) &&  0.670(20) &&  0.664(18) &&  0.665(21)  \\ 
                       && 5 &&  0.590(37) &&  0.610(42) &&  0.616(42) &&  0.625(26) &&  0.625(29) &&  0.636(18)  \\ 
\hline
$g_\perp$              && 1 &&  0.834(10) &&  0.820(11) &&  0.820(12) &&  0.828(13) &&  0.8246(93) &&  0.825(11)  \\ 
                       && 2 &&  0.719(11) &&  0.720(11) &&  0.719(12) &&  0.723(14) &&  0.730(15) &&  0.732(11)  \\ 
                       && 3 &&  0.640(12) &&  0.647(12) &&  0.645(12) &&  0.644(15) &&  0.656(14) &&  0.664(11)  \\ 
                       && 4 &&  0.581(25) &&  0.586(19) &&  0.578(23) &&  0.615(23) &&  0.600(15) &&  0.598(21)  \\ 
                       && 5 &&  0.508(22) &&  0.520(19) &&  0.514(22) &&  0.553(20) &&  0.543(16) &&  0.554(18)  \\ 
\hline
$g_+$                  && 1 &&  0.818(14) &&  0.806(10) &&  0.807(11) &&  0.8130(96) &&  0.8121(76) &&  0.8149(97)  \\ 
                       && 2 &&  0.696(13) &&  0.6977(97) &&  0.700(10) &&  0.7043(87) &&  0.7123(74) &&  0.7185(83)  \\ 
                       && 3 &&  0.611(15) &&  0.6128(95) &&  0.620(10) &&  0.6260(82) &&  0.6383(61) &&  0.6510(81)  \\ 
                       && 4 &&  0.547(19) &&  0.550(16) &&  0.550(18) &&  0.585(20) &&  0.572(14) &&  0.580(16)  \\ 
                       && 5 &&  0.477(28) &&  0.481(28) &&  0.487(29) &&  0.541(16) &&  0.527(17) &&  0.539(18)  \\ 
\hline
$g_0$                  && 1 &&  0.988(23) &&  0.985(24) &&  0.982(25) &&  1.009(26) &&  0.996(23) &&  0.990(22)  \\ 
                       && 2 &&  0.789(19) &&  0.788(14) &&  0.790(14) &&  0.808(13) &&  0.808(10) &&  0.814(12)  \\ 
                       && 3 &&  0.647(22) &&  0.647(13) &&  0.655(14) &&  0.669(11) &&  0.6775(93) &&  0.690(12)  \\ 
                       && 4 &&  0.550(22) &&  0.553(18) &&  0.552(19) &&  0.592(20) &&  0.577(16) &&  0.581(16)  \\ 
                       && 5 &&  0.461(25) &&  0.466(25) &&  0.468(26) &&  0.521(18) &&  0.507(15) &&  0.514(16)  \\ 
\hline
$h_\perp$              && 1 &&  0.859(23) &&  0.855(33) &&  0.858(24) &&  0.843(16) &&  0.8474(89) &&  0.852(13)  \\ 
                       && 2 &&  0.716(16) &&  0.713(17) &&  0.713(16) &&  0.697(20) &&  0.704(15) &&  0.718(14)  \\ 
                       && 3 &&  0.591(33) &&  0.590(21) &&  0.595(20) &&  0.579(26) &&  0.595(17) &&  0.613(14)  \\ 
                       && 4 &&  0.524(18) &&  0.530(18) &&  0.531(16) &&  0.533(19) &&  0.529(15) &&  0.531(21)  \\ 
                       && 5 &&  0.451(17) &&  0.463(18) &&  0.467(16) &&  0.476(17) &&  0.475(14) &&  0.479(14)  \\ 
\hline
$h_+$                  && 1 &&  1.638(56) &&  1.656(49) &&  1.700(45) &&  1.590(50) &&  1.609(42) &&  1.677(40)  \\ 
                       && 2 &&  1.387(39) &&  1.394(32) &&  1.432(28) &&  1.357(38) &&  1.379(30) &&  1.440(27)  \\ 
                       && 3 &&  1.166(70) &&  1.164(54) &&  1.198(46) &&  1.113(47) &&  1.154(36) &&  1.232(28)  \\ 
                       && 4 &&  0.991(36) &&  1.014(28) &&  1.044(26) &&  1.010(32) &&  1.023(25) &&  1.074(25)  \\ 
                       && 5 &&  0.876(35) &&  0.892(26) &&  0.924(25) &&  0.905(30) &&  0.919(24) &&  0.970(25)  \\ 
\hline
$\widetilde{h}_\perp$  && 1 &&  0.751(31) &&  0.744(22) &&  0.750(20) &&  0.739(18) &&  0.737(16) &&  0.743(18)  \\ 
                       && 2 &&  0.639(32) &&  0.642(23) &&  0.646(21) &&  0.624(19) &&  0.633(17) &&  0.649(18)  \\ 
                       && 3 &&  0.553(34) &&  0.558(25) &&  0.565(23) &&  0.541(31) &&  0.558(21) &&  0.578(20)  \\ 
                       && 4 &&  0.503(41) &&  0.512(29) &&  0.517(27) &&  0.524(24) &&  0.517(22) &&  0.526(26)  \\ 
                       && 5 &&  0.432(57) &&  0.446(45) &&  0.454(41) &&  0.467(37) &&  0.463(34) &&  0.485(32)  \\ 
\hline
$\widetilde{h}_+$      && 1 &&  0.762(31) &&  0.755(22) &&  0.761(21) &&  0.754(17) &&  0.749(15) &&  0.753(19)  \\ 
                       && 2 &&  0.656(31) &&  0.660(23) &&  0.661(21) &&  0.644(19) &&  0.651(16) &&  0.662(20)  \\ 
                       && 3 &&  0.567(34) &&  0.578(26) &&  0.583(23) &&  0.559(32) &&  0.576(23) &&  0.591(21)  \\ 
                       && 4 &&  0.525(42) &&  0.538(29) &&  0.537(27) &&  0.547(25) &&  0.537(22) &&  0.543(27)  \\ 
                       && 5 &&  0.456(48) &&  0.479(38) &&  0.479(36) &&  0.496(29) &&  0.488(28) &&  0.505(30)  \\ 
\hline\hline
\end{tabular}
\normalsize
\caption{\label{tab:FFlat}Values of the $\Lambda_c \to N$ form factors extracted for each nucleon momentum and each data set.}
\end{table}

\FloatBarrier

\providecommand{\href}[2]{#2}\begingroup\raggedright\endgroup


\begin{thebibliography}{10}

\bibitem{Ivanov:1996fj}
M.~A. Ivanov, V.~E. Lyubovitskij, J.~G. Körner, and P.~Kroll, ``{Heavy baryon
  transitions in a relativistic three quark model},''
  \href{http://dx.doi.org/10.1103/PhysRevD.56.348}{Phys. Rev. {\bfseries D56}
  (1997) 348--364},
\href{http://arxiv.org/abs/hep-ph/9612463}{{\ttfamily arXiv:hep-ph/9612463
  [hep-ph]}}.

\bibitem{Pervin:2005ve}
M.~Pervin, W.~Roberts, and S.~Capstick, ``{Semileptonic decays of heavy
  $\Lambda$ baryons in a quark model},''
  \href{http://dx.doi.org/10.1103/PhysRevC.72.035201}{Phys. Rev. {\bfseries
  C72} (2005) 035201},
\href{http://arxiv.org/abs/nucl-th/0503030}{{\ttfamily arXiv:nucl-th/0503030
  [nucl-th]}}.

\bibitem{Azizi:2009wn}
K.~Azizi, M.~Bayar, Y.~Sarac, and H.~Sundu, ``{Semileptonic $\Lambda_{b,c}$ to
  Nucleon Transitions in Full QCD at Light Cone},''
  \href{http://dx.doi.org/10.1103/PhysRevD.80.096007}{Phys. Rev. {\bfseries
  D80} (2009) 096007},
\href{http://arxiv.org/abs/0908.1758}{{\ttfamily arXiv:0908.1758 [hep-ph]}}.

\bibitem{Khodjamirian:2011jp}
A.~Khodjamirian, C.~Klein, T.~Mannel, and Y.-M. Wang, ``{Form Factors and
  Strong Couplings of Heavy Baryons from QCD Light-Cone Sum Rules},''
  \href{http://dx.doi.org/10.1007/JHEP09(2011)106}{JHEP {\bfseries 09} (2011)
  106},
\href{http://arxiv.org/abs/1108.2971}{{\ttfamily arXiv:1108.2971 [hep-ph]}}.

\bibitem{Gutsche:2014zna}
T.~Gutsche, M.~A. Ivanov, J.~G. Körner, V.~E. Lyubovitskij, and P.~Santorelli,
  ``{Heavy-to-light semileptonic decays of $\Lambda_b$ and $\Lambda_c$ baryons
  in the covariant confined quark model},''
  \href{http://dx.doi.org/10.1103/PhysRevD.90.114033,
  10.1103/PhysRevD.94.059902}{Phys. Rev. {\bfseries D90} no.~11, (2014)
  114033}, \href{http://arxiv.org/abs/1410.6043}{{\ttfamily arXiv:1410.6043
  [hep-ph]}}.
[Erratum: Phys. Rev.D94,no.5,059902(2016)].

\bibitem{Lu:2016ogy}
C.-D. Lü, W.~Wang, and F.-S. Yu, ``{Test flavor SU(3) symmetry in exclusive
  $\Lambda_c$ decays},''
  \href{http://dx.doi.org/10.1103/PhysRevD.93.056008}{Phys. Rev. {\bfseries
  D93} no.~5, (2016) 056008},
\href{http://arxiv.org/abs/1601.04241}{{\ttfamily arXiv:1601.04241 [hep-ph]}}.

\bibitem{Faustov:2016yza}
R.~N. Faustov and V.~O. Galkin, ``{Semileptonic decays of $\Lambda _c$ baryons
  in the relativistic quark model},''
  \href{http://dx.doi.org/10.1140/epjc/s10052-016-4492-z}{Eur. Phys. J.
  {\bfseries C76} no.~11, (2016) 628},
\href{http://arxiv.org/abs/1610.00957}{{\ttfamily arXiv:1610.00957 [hep-ph]}}.

\bibitem{Li:2016qai}
C.-F. Li, Y.-L. Liu, K.~Liu, C.-Y. Cui, and M.-Q. Huang, ``{Analysis of the
  semileptonic decay ${{\rm{\Lambda }}}_{c} \rightarrow {{ne}}^{+}{\nu
  }_{e}$},'' \href{http://dx.doi.org/10.1088/1361-6471/aa68f1}{J. Phys.
  {\bfseries G44} no.~7, (2017) 075006},
\href{http://arxiv.org/abs/1610.05418}{{\ttfamily arXiv:1610.05418 [hep-ph]}}.

\bibitem{deBoer:2015boa}
S.~de~Boer and G.~Hiller, ``{Flavor and new physics opportunities with rare
  charm decays into leptons},''
  \href{http://dx.doi.org/10.1103/PhysRevD.93.074001}{Phys. Rev. {\bfseries
  D93} no.~7, (2016) 074001},
\href{http://arxiv.org/abs/1510.00311}{{\ttfamily arXiv:1510.00311 [hep-ph]}}.

\bibitem{Fajfer:2015mia}
S.~Fajfer and N.~Košnik, ``{Prospects of discovering new physics in rare charm
  decays},'' \href{http://dx.doi.org/10.1140/epjc/s10052-015-3801-2}{Eur. Phys.
  J. {\bfseries C75} no.~12, (2015) 567},
\href{http://arxiv.org/abs/1510.00965}{{\ttfamily arXiv:1510.00965 [hep-ph]}}.

\bibitem{Feldmann:2017izn}
T.~Feldmann, B.~Müller, and D.~Seidel, ``{$D \to \rho \,\ell^+\ell^-$ decays
  in the QCD factorization approach},''
  \href{http://dx.doi.org/10.1007/JHEP08(2017)105}{JHEP {\bfseries 08} (2017)
  105},
\href{http://arxiv.org/abs/1705.05891}{{\ttfamily arXiv:1705.05891 [hep-ph]}}.

\bibitem{Aaij:2017nsd}
{\bfseries LHCb} Collaboration, R.~Aaij {\em et~al.}, ``{Search for the rare
  decay $\Lambda_{c}^{+} \to p\mu^+\mu^-$},''
\href{http://arxiv.org/abs/1712.07938}{{\ttfamily arXiv:1712.07938 [hep-ex]}}.

\bibitem{Lees:2011hb}
{\bfseries BaBar} Collaboration, J.~P. Lees {\em et~al.}, ``{Searches for Rare
  or Forbidden Semileptonic Charm Decays},''
  \href{http://dx.doi.org/10.1103/PhysRevD.84.072006}{Phys. Rev. {\bfseries
  D84} (2011) 072006},
\href{http://arxiv.org/abs/1107.4465}{{\ttfamily arXiv:1107.4465 [hep-ex]}}.

\bibitem{Kodama:1995ia}
{\bfseries E653} Collaboration, K.~Kodama {\em et~al.}, ``{Upper limits of
  charm hadron decays to two muons plus hadrons},''
\href{http://dx.doi.org/10.1016/0370-2693(94)01610-O}{Phys. Lett. {\bfseries
  B345} (1995) 85--92}.

\bibitem{Feldmann:2011xf}
T.~Feldmann and M.~W.~Y. Yip, ``{Form Factors for $\Lambda_b \to \Lambda$
  Transitions in {SCET}},'' \href{http://dx.doi.org/10.1103/PhysRevD.85.014035,
  10.1103/PhysRevD.86.079901}{Phys. Rev. {\bfseries D85} (2012) 014035},
  \href{http://arxiv.org/abs/1111.1844}{{\ttfamily arXiv:1111.1844 [hep-ph]}}.
[Erratum: Phys. Rev.D86,079901(2012)].

\bibitem{Iwasaki:1984cj}
Y.~Iwasaki and T.~Yoshie, ``{Renormalization group improved action for $SU(3)$
  lattice gauge theory and the string tension},''
\href{http://dx.doi.org/10.1016/0370-2693(84)91500-4}{Phys.Lett. {\bfseries
  B143} (1984) 449}.

\bibitem{Kaplan:1992bt}
D.~B. Kaplan, ``{A Method for simulating chiral fermions on the lattice},''
  \href{http://dx.doi.org/10.1016/0370-2693(92)91112-M}{Phys.Lett. {\bfseries
  B288} (1992) 342--347},
\href{http://arxiv.org/abs/hep-lat/9206013}{{\ttfamily arXiv:hep-lat/9206013}}.

\bibitem{Furman:1994ky}
V.~Furman and Y.~Shamir, ``{Axial symmetries in lattice QCD with Kaplan
  fermions},'' \href{http://dx.doi.org/10.1016/0550-3213(95)00031-M}{Nucl.Phys.
  {\bfseries B439} (1995) 54--78},
\href{http://arxiv.org/abs/hep-lat/9405004}{{\ttfamily arXiv:hep-lat/9405004}}.

\bibitem{Shamir:1993zy}
Y.~Shamir, ``{Chiral fermions from lattice boundaries},''
  \href{http://dx.doi.org/10.1016/0550-3213(93)90162-I}{Nucl.Phys. {\bfseries
  B406} (1993) 90--106},
\href{http://arxiv.org/abs/hep-lat/9303005}{{\ttfamily arXiv:hep-lat/9303005}}.

\bibitem{Brown:2014ena}
Z.~S. Brown, W.~Detmold, S.~Meinel, and K.~Orginos, ``{Charmed bottom baryon
  spectroscopy from lattice QCD},''
  \href{http://dx.doi.org/10.1103/PhysRevD.90.094507}{Phys.Rev. {\bfseries D90}
  (2014) 094507},
\href{http://arxiv.org/abs/1409.0497}{{\ttfamily arXiv:1409.0497 [hep-lat]}}.

\bibitem{Aoki:2010dy}
{\bfseries RBC, UKQCD} Collaboration, Y.~Aoki {\em et~al.}, ``{Continuum Limit
  Physics from 2+1 Flavor Domain Wall QCD},''
  \href{http://dx.doi.org/10.1103/PhysRevD.83.074508}{Phys. Rev. {\bfseries
  D83} (2011) 074508},
\href{http://arxiv.org/abs/1011.0892}{{\ttfamily arXiv:1011.0892 [hep-lat]}}.

\bibitem{Detmold:2015aaa}
W.~Detmold, C.~Lehner, and S.~Meinel, ``{$\Lambda_b \to p \ell^-
  \bar{\nu}_\ell$ and $\Lambda_b \to \Lambda_c \ell^- \bar{\nu}_\ell$ form
  factors from lattice QCD with relativistic heavy quarks},''
  \href{http://dx.doi.org/10.1103/PhysRevD.92.034503}{Phys. Rev. {\bfseries
  D92} no.~3, (2015) 034503},
\href{http://arxiv.org/abs/1503.01421}{{\ttfamily arXiv:1503.01421 [hep-lat]}}.

\bibitem{Meinel:2010pv}
S.~Meinel, ``{Bottomonium spectrum at order $v^6$ from domain-wall lattice QCD:
  Precise results for hyperfine splittings},''
  \href{http://dx.doi.org/10.1103/PhysRevD.82.114502}{Phys. Rev. {\bfseries
  D82} (2010) 114502},
\href{http://arxiv.org/abs/1007.3966}{{\ttfamily arXiv:1007.3966 [hep-lat]}}.

\bibitem{Hashimoto:1999yp}
S.~Hashimoto, A.~X. El-Khadra, A.~S. Kronfeld, P.~B. Mackenzie, S.~M. Ryan,
  {\em et~al.}, ``{Lattice QCD calculation of $\bar{B} \to D \ell \bar{\nu}$
  decay form-factors at zero recoil},''
  \href{http://dx.doi.org/10.1103/PhysRevD.61.014502}{Phys.Rev. {\bfseries D61}
  (1999) 014502},
\href{http://arxiv.org/abs/hep-ph/9906376}{{\ttfamily arXiv:hep-ph/9906376
  [hep-ph]}}.

\bibitem{ElKhadra:2001rv}
A.~X. El-Khadra, A.~S. Kronfeld, P.~B. Mackenzie, S.~M. Ryan, and J.~N. Simone,
  ``{The Semileptonic decays $B \to \pi \ell \nu$ and $D \to \pi \ell \nu$ from
  lattice QCD},'' \href{http://dx.doi.org/10.1103/PhysRevD.64.014502}{Phys.Rev.
  {\bfseries D64} (2001) 014502},
\href{http://arxiv.org/abs/hep-ph/0101023}{{\ttfamily arXiv:hep-ph/0101023
  [hep-ph]}}.

\bibitem{Lehner:2012bt}
C.~Lehner, ``{Automated lattice perturbation theory and relativistic heavy
  quarks in the Columbia formulation},'' PoS {\bfseries LATTICE2012} (2012)
  126,
\href{http://arxiv.org/abs/1211.4013}{{\ttfamily arXiv:1211.4013 [hep-lat]}}.

\bibitem{Lehnercharmlight}
C.~Lehner. Private communication, 2016.

\bibitem{Detmold:2016pkz}
W.~Detmold and S.~Meinel, ``{$\Lambda_b \to \Lambda \ell^+ \ell^-$ form
  factors, differential branching fraction, and angular observables from
  lattice QCD with relativistic $b$ quarks},''
  \href{http://dx.doi.org/10.1103/PhysRevD.93.074501}{Phys. Rev. {\bfseries
  D93} no.~7, (2016) 074501},
\href{http://arxiv.org/abs/1602.01399}{{\ttfamily arXiv:1602.01399 [hep-lat]}}.

\bibitem{ElKhadra:1996mp}
A.~X. El-Khadra, A.~S. Kronfeld, and P.~B. Mackenzie, ``{Massive fermions in
  lattice gauge theory},''
  \href{http://dx.doi.org/10.1103/PhysRevD.55.3933}{Phys. Rev. {\bfseries D55}
  (1997) 3933--3957},
\href{http://arxiv.org/abs/hep-lat/9604004}{{\ttfamily arXiv:hep-lat/9604004
  [hep-lat]}}.

\bibitem{Blum:2014tka}
{\bfseries RBC, UKQCD} Collaboration, T.~Blum {\em et~al.}, ``{Domain wall QCD
  with physical quark masses},''
  \href{http://dx.doi.org/10.1103/PhysRevD.93.074505}{Phys. Rev. {\bfseries
  D93} no.~7, (2016) 074505},
\href{http://arxiv.org/abs/1411.7017}{{\ttfamily arXiv:1411.7017 [hep-lat]}}.

\bibitem{Meinel:2016dqj}
S.~Meinel, ``{$\Lambda_c \to \Lambda l^+ \nu_l$ form factors and decay rates
  from lattice QCD with physical quark masses},''
  \href{http://dx.doi.org/10.1103/PhysRevLett.118.082001}{Phys. Rev. Lett.
  {\bfseries 118} no.~8, (2017) 082001},
\href{http://arxiv.org/abs/1611.09696}{{\ttfamily arXiv:1611.09696 [hep-lat]}}.

\bibitem{Bourrely:2008za}
C.~Bourrely, I.~Caprini, and L.~Lellouch, ``{Model-independent description of
  $B \to \pi \ell \nu$ decays and a determination of $|V_{ub}|$},''
  \href{http://dx.doi.org/10.1103/PhysRevD.82.099902,
  10.1103/PhysRevD.79.013008}{Phys.Rev. {\bfseries D79} (2009) 013008},
\href{http://arxiv.org/abs/0807.2722}{{\ttfamily arXiv:0807.2722 [hep-ph]}}.

\bibitem{Bailey:2015dka}
J.~A. Bailey {\em et~al.}, ``{$B\to Kl^+l^-$ decay form factors from
  three-flavor lattice QCD},''
  \href{http://dx.doi.org/10.1103/PhysRevD.93.025026}{Phys. Rev. {\bfseries
  D93} no.~2, (2016) 025026},
\href{http://arxiv.org/abs/1509.06235}{{\ttfamily arXiv:1509.06235 [hep-lat]}}.

\bibitem{Patrignani:2016xqp}
{\bfseries Particle Data Group} Collaboration, C.~Patrignani {\em et~al.},
  ``{Review of Particle Physics},''
\href{http://dx.doi.org/10.1088/1674-1137/40/10/100001}{Chin. Phys. {\bfseries
  C40} no.~10, (2016) 100001}.

\bibitem{supplementalmaterial}
See Supplemental Material at \url{https://arxiv.org/src/1712.05783/anc}
for files containing the form factor parameter values and covariances.

\bibitem{UTfit}
{\bfseries UTfit} Collaboration, ``Standard model fit results: Summer 2016.''
\newblock \url{http://www.utfit.org/UTfit/ResultsSummer2016SM}.

\bibitem{Ablikim:2015prg}
{\bfseries BESIII} Collaboration, M.~Ablikim {\em et~al.}, ``{Measurement of
  the absolute branching fraction for $\Lambda^+_{c}\to \Lambda e^+\nu_e$},''
  \href{http://dx.doi.org/10.1103/PhysRevLett.115.221805}{Phys. Rev. Lett.
  {\bfseries 115} no.~22, (2015) 221805},
\href{http://arxiv.org/abs/1510.02610}{{\ttfamily arXiv:1510.02610 [hep-ex]}}.

\bibitem{deBoer:2016dcg}
S.~de~Boer, B.~Müller, and D.~Seidel, ``{Higher-order Wilson coefficients for
  $c \to u$ transitions in the standard model},''
  \href{http://dx.doi.org/10.1007/JHEP08(2016)091}{JHEP {\bfseries 08} (2016)
  091},
\href{http://arxiv.org/abs/1606.05521}{{\ttfamily arXiv:1606.05521 [hep-ph]}}.

\bibitem{deBoerphdthesis}
S.~de~Boer, \href{http://dx.doi.org/10.17877/DE290R-18060}{{\em {Probing the
  standard model with rare charm decays}}}.
\newblock PhD thesis, {Technische Universität Dortmund}, 2017.
\newblock \url{{http://hdl.handle.net/2003/36043}}.

\bibitem{deBoer:2017way}
S.~de~Boer, ``{Two loop virtual corrections to $b\rightarrow (d,s)\ell ^+\ell
  ^-$ and $c\rightarrow u\ell ^+\ell ^-$ for arbitrary momentum transfer},''
  \href{http://dx.doi.org/10.1140/epjc/s10052-017-5364-x}{Eur. Phys. J.
  {\bfseries C77} no.~11, (2017) 801},
\href{http://arxiv.org/abs/1707.00988}{{\ttfamily arXiv:1707.00988 [hep-ph]}}.

\bibitem{deBoerWC}
S.~de~Boer. Private communication, 2017.

\bibitem{Chetyrkin:2000yt}
K.~G. Chetyrkin, J.~H. Kühn, and M.~Steinhauser, ``{RunDec: A Mathematica
  package for running and decoupling of the strong coupling and quark
  masses},'' \href{http://dx.doi.org/10.1016/S0010-4655(00)00155-7}{Comput.
  Phys. Commun. {\bfseries 133} (2000) 43--65},
\href{http://arxiv.org/abs/hep-ph/0004189}{{\ttfamily arXiv:hep-ph/0004189
  [hep-ph]}}.

\bibitem{Fajfer:2005ke}
S.~Fajfer and S.~Prelovsek, ``{Effects of littlest Higgs model in rare $D$
  meson decays},'' \href{http://dx.doi.org/10.1103/PhysRevD.73.054026}{Phys.
  Rev. {\bfseries D73} (2006) 054026},
\href{http://arxiv.org/abs/hep-ph/0511048}{{\ttfamily arXiv:hep-ph/0511048
  [hep-ph]}}.

\bibitem{Boer:2014kda}
P.~B$\ddot{\mathrm{o}}$er, T.~Feldmann, and D.~van Dyk, ``{Angular Analysis of
  the Decay $\Lambda_b \to \Lambda (\to N \pi) \ell^+\ell^-$},''
  \href{http://dx.doi.org/10.1007/JHEP01(2015)155}{JHEP {\bfseries 01} (2015)
  155},
\href{http://arxiv.org/abs/1410.2115}{{\ttfamily arXiv:1410.2115 [hep-ph]}}.

\bibitem{Gutsche:2013pp}
T.~Gutsche, M.~A. Ivanov, J.~G. K$\ddot{\mathrm{o}}$rner, V.~E. Lyubovitskij,
  and P.~Santorelli, ``{Rare baryon decays $\Lambda_b \to \Lambda {l^{+}l^{-}}
  (l=e, \mu, \tau)$ and $\Lambda_b \to \Lambda\gamma$ : differential and total
  rates, lepton- and hadron-side forward-backward asymmetries},''
  \href{http://dx.doi.org/10.1103/PhysRevD.87.074031}{Phys.Rev. {\bfseries D87}
  (2013) 074031},
\href{http://arxiv.org/abs/1301.3737}{{\ttfamily arXiv:1301.3737 [hep-ph]}}.

\bibitem{Broadhurst:1994se}
D.~J. Broadhurst and A.~G. Grozin, ``{Matching QCD and HQET heavy - light
  currents at two loops and beyond},''
  \href{http://dx.doi.org/10.1103/PhysRevD.52.4082}{Phys. Rev. {\bfseries D52}
  (1995) 4082--4098},
\href{http://arxiv.org/abs/hep-ph/9410240}{{\ttfamily arXiv:hep-ph/9410240
  [hep-ph]}}.

\bibitem{Azizi:2010zzb}
K.~Azizi, M.~Bayar, Y.~Sarac, and H.~Sundu, ``{FCNC transitions of
  $\Lambda_{b,c}$ to nucleon in SM},''
\href{http://dx.doi.org/10.1088/0954-3899/37/11/115007}{J. Phys. {\bfseries
  G37} (2010) 115007}.

\bibitem{Sirvanli:2016wnr}
B.~B. \c{S}irvanli, ``{Search for $c \to ul^+l^-$ transition in charmed baryon
  decays},''
\href{http://dx.doi.org/10.1103/PhysRevD.93.034027}{Phys. Rev. {\bfseries D93}
  no.~3, (2016) 034027}.

\bibitem{Paul:2011ar}
A.~Paul, I.~I. Bigi, and S.~Recksiegel, ``{On $D\to X_u l^+ l^-$ within the
  Standard Model and Frameworks like the Littlest Higgs Model with T Parity},''
  \href{http://dx.doi.org/10.1103/PhysRevD.83.114006}{Phys. Rev. {\bfseries
  D83} (2011) 114006},
\href{http://arxiv.org/abs/1101.6053}{{\ttfamily arXiv:1101.6053 [hep-ph]}}.

\bibitem{Edwards:2004sx}
{\bfseries SciDAC, LHPC, UKQCD} Collaboration, R.~G. Edwards and B.~Joo, ``{The
  Chroma software system for lattice QCD},''
  \href{http://dx.doi.org/10.1016/j.nuclphysbps.2004.11.254}{Nucl.Phys.Proc.Suppl.
  {\bfseries 140} (2005) 832},
\href{http://arxiv.org/abs/hep-lat/0409003}{{\ttfamily arXiv:hep-lat/0409003
  [hep-lat]}}.

\end{thebibliography}
\end{document}